\documentstyle[12pt]{article}
\newcommand{\be}{\begin{equation}}
\newcommand{\ee}{\end{equation}}
\newcommand{\ba}{\begin{eqnarray}}
\newcommand{\ea}{\end{eqnarray}}

\begin{document}

\title{{\bf Loop quantum gravity and light propagation}}
\author{{\rm Jorge Alfaro$^a$\thanks{%
jalfaro@fis.puc.cl}, Hugo A. Morales-T\'ecotl$^{bc}$\thanks{
hugo@xanum.uam.mx} \thanks{%
Associate member of Abdus Salam International Centre for Theoretical
Physics, Trieste, Italy.}} \\
and Luis F. Urrutia$^d$\thanks{%
urrutia@nuclecu.unam.mx} \\
\\
$^a$Facultad de F\'{\i}sica \\
Pontificia Universidad Cat\'olica de Chile\\
Casilla 306, Santiago 22, Chile\\
$^b$Center for Gravitational Physics and Geometry, Department of Physics \\
The Pennsylvania State University, 104 Davey Lab\\
University Park PA 16802, USA\\
$^c$Departamento de F\'{\i}sica \\
Universidad Aut\'onoma Metropolitana Iztapalapa \\
A.P. 55-534, M\'exico D.F. 09340, M\'exico\\
$^d$Departamento de F\'{\i}sica de Altas Energ\'{\i}as\\
Instituto de Ciencias Nucleares \\
Universidad Nacional Aut\'onoma de M\'exico \\
A.P. 70-543, M\'exico D.F. 04510, M\'exico }
\maketitle

\begin{abstract}
Within loop quantum gravity we construct a
coarse-grained approximation for the Einstein-Maxwell theory  that yields effective Maxwell equations in flat
spacetime comprising Planck scale corrections.

The corresponding Hamiltonian is defined as the expectation value of the
electromagnetic term in the Einstein-Maxwell Hamiltonian constraint, regularized a la
Thiemann, with respect to a would-be semiclassical state. The resulting energy
dispersion relations entail Planck scale corrections to those in flat spacetime. Both
the helicity dependent contribution of Gambini and Pullin \cite{GP} and, for a value of
a parameter of our approximation, that of Ellis et. al. \cite{ELLISETAL} are recovered.
The electric/magnetic asymmetry in the regularization procedure yields nonlinearities
only in the magnetic sector which are briefly discussed. Observations of cosmological
Gamma Ray Bursts might eventually lead to the needed accuracy to study some of these
quantum gravity effects.

\end{abstract}

\clearpage

\vfill \eject

\baselineskip=20pt

\section{Introduction}

It has been recently suggested that quantum gravity effects might be indeed observable
\cite{AC}-\cite{AHLU}. Among the diverse phenomena  where these observations could take place we can distinguish (see e.g. \cite{ACPol}-\cite{ACEssay}): (i) strain noise induced in gravitational
wave detectors, (ii) neutral Kaon systems and (iii) energy dependent time of arrival of
photons/neutrinos from distant sources. Experimental sensitivities in each of the above situations can be argued to be at the edge of what is required to reveal quantum gravity phenomena, namely Planck
length sensitivities. In this work we focus on the third possibility. The idea is to look for
modified dispersion relations of photons with energy $E$ and momentum $\vec k $, of
the form
\begin{eqnarray}
c^2 \, {\vec k}{\,\,}^2= E^2\left(1 + \xi\, \frac{E}{E_{QG}}+ {\cal O}\left(%
\frac{E}{E_{QG}}\right)^2 \right)\,,\quad  \label{eq:drel}
\end{eqnarray}
where $\xi$ is a numerical factor of order one and $E_{QG}$ is an energy scale of order $\leq 10^{19}$ GeV, which signals the need of considering the quantum character of gravity.
The above expression leads to the following modification of the speed of light {\em in
vacuum}
\begin{eqnarray}
v= \frac{\partial E}{\partial k}= c \left(1- \xi\, \frac{E}{E_{QG}}+ {\cal O}%
\left(\frac{E}{E_{QG}}\right)^2\right)\,,
\end{eqnarray}
which implies a retardation time
\begin{eqnarray}
\Delta t \approx \xi\, \frac{E}{E_{QG}}\, \frac{L}{c}\,,
\label{deltat}
\end{eqnarray}
with respect to a signal
propagating with speed $c$. When we consider  cosmological distances $L\approx 10^{10}\,$ ly and an energy scale $E_{QG} \approx 10^{19}\, {\rm
GeV}$ in (\ref{deltat}) , the corresponding values are $\Delta t \approx 10^{-3}$ {\rm s} ($E\approx
20 {\rm MeV}$) and $\Delta t\approx 10^{-5}$ {\rm s} ($E\approx 0.20 {\rm MeV}$).
In order to detect such effects, an experimental time resolution $\delta t$ at least of order
$\Delta t$ is required. Thus, short and intense bursts traveling  large distances
would be the best candidates. Recent observations pointing towards the possibility of
attaining  such conditions are:
(i) some Gamma Ray Bursts (GRB) originate at cosmological dis\-tances, ($%
\sim 10^{10} \ {\rm l y})$ \cite{METZ} and  (ii) sensitivities $\, \delta t $
up to submillisecond scale have been achieved in GRB observations %
\cite{BHAT} and they  are expected to improve in future spatial experiments \cite%
{meszaros}.

It is thus timely to investigate whether candidate quantum gravity theories can
account for (\ref{eq:drel}). Modifications to Maxwell's equations in vacuum, induced
by quantum gravity effects, have been calculated by Gambini and Pullin \cite{GP}.
By considering  a semiclassical regime in which the electromagnetic field is a classical object
whereas space is described by loop quantum gravity, they obtained  the dispersion relations
\begin{eqnarray}
\omega_{\pm}(k)=k \left(1 \mp 2\, \xi\, \ell_P\, k \right)\;,\,\quad k=|\vec k|,
\end{eqnarray}
where $\pm$ labels  the helicity of the photon and $\ell_P\approx 10^{-33}\,{\rm
cm}$ is the Planck length. This modification, being helicity dependent, yields parity
violation and birefringent effects.

On the other hand, a string theory approach proposed by Ellis et al. \cite{ELLISETAL} suggests that a
D-brane recoil in the quantum-gravitational foam induces a distortion in the
surrounding space, which modifies the photon propagation properties. For a review of
this approach see for example Ref. \cite{REVIEW}. The dispersion
relations obtained via this second procedure are
\begin{eqnarray}
\omega(k)= k\left(1 - \xi\, \frac{k}{M_D}\right)\,,
\end{eqnarray}
with $\xi >0$. They arise from parity conserving corrections to Maxwell's equations
which lead to a first order ( in $1/M_D$) helicity independent effect in the dispersion
relations that is linear in the photon energy. No birefringent effects appear at this
order. In this approach, the red-shifted difference in the time arrival of two photons
with present-day energies $E_1$ and $E_2$ has been calculated. For the BATSE data, when
the redshifts $z$ of the GRB are known, a small subset of coincident photon pulses
corresponding to channel 1 (20-50\,keV) and channel 3 (100-300\,keV) are fitted and
$\Delta t$ is calculated \cite{ELLISETAL}. No significant effect in the data available
is found. On the other hand, none of the pulses studied exhibited microbursts having a
short time structure on scales $\leq 10^{-2}$ s. Were this the case, the sensitivity of
the analysis would be greatly improved. Alternative studies based upon effective
perturbative quantum gravity \cite{oneloop-effqg}, open system techniques
\cite{opensystems} and quantum light cone fluctuations \cite{Ford99} have been
performed.

The study of cosmological neutrinos could also provide an excellent ar\-e\-na to probe quantum
gravity induced propagation effects because space is practically transparent to them,
even at very high energies. In fact, the
fireball model, which is one of the  most popular  models of GRB, predicts the generation of $10^{14}-10^{19}$ eV Neutrino Bursts (NB)
\cite{WAX,VIETRI}. The planned Neutrino Burster Experiment (NuBE) will measure the
flux of ultra high energy neutrinos ($>$ 10 TeV) over a $\sim 1{\rm km}^2$ effective
area, in coincidence with satellite measured GRB's \cite{ROY}. It is expected to
detect $\approx$ 20 events per year, according to the fireball model. Other relevant
experiment aimed at observing ultra high energy cosmic rays, including neutrinos, is
the OWL/Airwatch project which expects to see $\sim
3\times 10 ^3-10^5$ cosmic ray events with energies $>10^{20}$eV \cite%
{cline,halzen}. Notably, this experiment is able to investigate time correlations
among high energy neutrinos and gamma-rays. Hence in the foreseeable future, it might
be possible to study quantum gravity effects on observed astrophysical neutrinos and
photons. At the least, such observations could be used to restrict quantum gravity
theories.

 Motivated by these
interesting possibilities, we have calculated the quantum gravity induced
modifications to neutrino propagation  in \cite{URRU1}, within the loop quantum gravity
framework. We obtained corrections to the velocity of propagation which are proportional
to $(k\, \ell_P)$ together with helicity dependent corrections of order $(k\, \ell_P)^2$.
The energy dependence in the first case coincides with that found later by Ellis et.
al. using string theory methods \cite{ELLISFERM}.

In this work we extend our approach to the case of photons. The corrections obtained
within our approximation contain those of \cite{GP} and, for a given value of a
parameter in our scheme, those of \cite{ELLISETAL}. Besides, we briefly
discuss  higher order nonlinearities arising in  the magnetic sector of the effective Maxwell
Hamiltonian.

 The organization of the paper is as
follows. In section \ref{LQG} we recall some basic aspects of loop quantum gravity
which are necessary for our analysis. After reviewing Thiemann's regularization of the
Hamiltonian constraint of the Einstein-Maxwell theory in section \ref{thiereg}, we provide a
general description of our approximation  in section \ref{GIC}. The
corrections arising from the electric and magnetic sectors are calculated in section
\ref{thieregM}. Once the effective Hamiltonian is obtained, we derive  the modified Maxwell
equations together with  the corresponding dispersion
relations  in section \ref{MMEQ}. Section \ref{NLT} contains a brief analysis of the non-linear
effects arising from the magnetic sector. An outline of red shift effects on
cosmological photon time delays is given in section \ref{SPHOTD}. Finally,
section \ref{disc} contains a discussion of the results.

\section{Loop quantum gravity}

\label{LQG} In this section we summarize the main ingredients defining this approach
also denominated quantum geometry \cite{volumeop}. Among the main results along this
approach one finds: i) well defined geometric operators possessing a discrete spectrum, thus
evidencing discreteness of space \cite{volumeop}, ii) a microscopic account for black
hole entropy \cite{CanonicalBlackHole} and, more recently, hints on quantum avoidance
of a would be classical cosmological singularity \cite{canonicalnosingular}. (For a
review on these topics see for example Ref. \cite{RROV}.)

 To begin with it is assumed that
the spacetime manifold $M$ has topology $\Sigma\times I\!\!R$, with $\Sigma$ a
Riemannian 3-manifold. Here a co-triad $e^i_a$ is defined, with $a,b,c, \dots $ being
spatial tensor indices and $i,j,k,\dots$ being $su(2)$ indices. Thus the
corresponding three-metric is given by $q_{ab}=e^i_ae^i_b$. In addition, a field
$K^i_a$ is defined by
$K_{ab}=sgn [\det(e^j_c)] K^i_a e^i_b$, which is related to the extrinsic curvature $K_{ab}$ of $\Sigma$ .
A canonical pair for the gravitational phase space is $(K^i_a, E_j^b/\kappa)$%
, where $E^a_i=\frac{1}{2}\epsilon^{abc}\epsilon_{ijk}e^j_be^k_c$ and $%
\kappa $ is Newton's constant. It turns out that such a canonical pair yields a
complicated form for the Hamiltonian constraint of general relativity. A convenient
canonical pair, making this constraint polynomial, was introduced by Ashtekar
\cite{Ash}. Nevertheless, two severe difficulties to proceed with the quantization
remained: (i) the implementation of a diffeomorphism covariant regularization for the
density-weight two Hamiltonian constraint hereby obtained and (ii) the extension to
non-compact groups of the diffeomorphism covariant techniques already developed for
gauge theories with compact groups \cite{qdiffgauge}. In fact, the Ashtekar variables
\, ($\, ^{I\!\!\!\!C}\!\!A^i_a=\Gamma^i_a-iK^i_a$, \,\,  $\, iE^a_i/\kappa$) \cite{Ash} , with
$\Gamma^i_a$ being the torsion free connection compatible with $e^i_a$, are complex
valued. Namely the gauge group is $SL(2,I\!\!\!\!C)$, which is non compact.

Some proposals to come to terms with difficulty (ii) were: to consider real
connection variables \cite{Barbero}, to implement a Wick transform %
\cite{WICK} and to define tractable reality constraints \cite{REALITY}. All of these
left open (i).  Thiemann subsequently  proposed to solve (i) and (ii) by incorporating real
connection variables while keeping the density weight one character of the
Hamiltonian constraint. He further provided a quantum version of the theory in the  pure
gravity case, as well as in those cases including the coupling  of  matter to gravity \cite{Thiemann}. His approach is next  reviewed,
since we rely upon it for our analysis of the electromagnetic case.

Let us start with the following canonical pairs for the Einstein-Maxwell
system: $(A^i_a=\Gamma^i_a+K^i_a, E^a_i/\kappa)$ for the gravity sector and $%
(\underline{A}_a, \underline E^a/Q^2)$ for the electromagnetic sector. The
latter has gauge group $U(1)$ and $Q$ is the electromagnetic coupling
constant, related to the dimensionless fine structure constant by $%
\alpha_{EM}=Q^2\hbar$. The corresponding contributions to the Hamiltonian
constraint are
\begin{eqnarray}  \label{hmax}
H_{{\rm Einstein}}[N] &=& \int_{\Sigma} d^3x \; N
\frac{1}{\kappa\sqrt{\det q}} {\rm tr} \left( 2\left [%
K_a,K_b]-F_{ab}\right) [E_a,E_b] \right), \nonumber \\
H_{{\rm Maxwell}}[N] &=& \int_{\Sigma} d^3x \; N \frac{q_{ab}}{2Q^2\sqrt{\det q}}
[\underline E ^a \underline E^b + \underline B ^a \underline B^b].
\end{eqnarray}
Here $F_{ab}$ is the curvature of $A_a$ and $\underline B^b$ is the magnetic field of
the $U(1)$ connection $\underline A$. The actual classical configuration space is the space ${\cal
A}/{\cal G}$ of (both) connections modulo their gauge transformations. Indeed, this is
what occurs in gauge theories where the fundamental field is a connection. This
completes the classical description of the phase space of the theory.

The quantum arena is given as follows \cite{qdiffgauge}. As in any quantum field
theory, because of the infinite number of degrees of freedom, an enlargement of the
classical configuration space is required. This is far from trivial since the
measures defining the scalar product, which are required to provide a Hilbert space, get
concentrated on distributional fields and hence lie outside the classical configuration
space. The key idea to build up such an enlargement is to make Wilson loop variables
(traces of parallel transport operators) well defined. The
resulting space $\overline{{\cal A}/{\cal G }}$ can be thought of as the limit of
configuration spaces of lattice gauge theories for all possible {\em floating} (i.e. not necessarily rectangular) lattices. Hence, geometric structures on lattice
configuration space are used to implement a geometric structure on $\overline{{\cal
A}/{\cal G}}$. This enables to define a background independent calculus on
it which, in turn, leads to the construction of the relevant measures, the Hilbert space and the regulated operators.

In line with the Dirac procedure for constrained systems, one first ignores
the constraints and constructs an auxiliary Hilbert space ${\cal H}_{{\rm aux%
}}$, so that the set of elementary real functions on the full phase space is
represented by self-adjoint operators in ${\cal H}_{{\rm aux}}$. It turns
out that ${\cal H}_{{\rm aux}}$ is just $L^2(\overline{{\cal A}/{\cal G}}%
,\mu_0 )$, with $\mu_0$ being a suitable measure that implements the self-adjointness property.

Diffeomorphism constraints are well defined operators on ${\cal H}_{{\rm aux}%
}$ yielding no anomalies. There exists a dense subspace $\Phi$ of ${\cal H}_{%
{\rm aux}}$ so that its topological dual $\Phi^{\prime}$ includes a complete
set of solutions to the diffeomorphism constraint. They are characterized by
generalized knots (i.e. diffeomorphism invariant classes of graphs). Besides,
a diffeomorphism invariant Hilbert space is obtained for such states with an
inner product that represents real observables by self-adjoint operators.

Furthermore, ${\cal H}_{{\rm aux}}$ admits a basis in terms of the so called
spin network states. A spin network is a triple $(\alpha,\vec \jmath, \vec
p) $ consisting of a graph $\alpha$, a {\it coloring} defined by a  set of irreducible representations $%
(\jmath_1,\dots,\jmath_n)$ of $SU(2)$, with $\jmath_i$ corresponding to the edge
$e_i$ of $\alpha$ and a set of contractors $(p_1,\dots,p_m)$. Here, a contractor $p_k$
is just an intertwining operator from the tensor product of representations of the
incoming edges at the vertex $v_k$ to the tensor product of representations of the
outgoing edges. Compactness of $SU(2)$ makes the vector space of all possible
contractors $p_k$ finite, for a given $\vec \jmath$ and vertex $v_k$. An additional
non-degeneracy condition is included: $j_e$ is not trivial for any edge $e$ and
$\alpha$ is taken to be
minimal (i.e. any other $\alpha^{\prime}$, occupying the same points in $%
\Sigma$ as $\alpha$, can always be built by subdividing the edges of $\alpha$%
, but not the other way around).

A spin network state is a $C^{\infty}$ cylindrical function (a function that depends
on the connection at the finite number of edges of a graph) on $\overline{{\cal
A}/{\cal G}}$, constructed from a spin network
\begin{equation}
T_{\alpha,{\vec \jmath},{\vec c}} [A]:=
[\otimes_{i=1}^n\jmath_i(h_{e_i}(A))\bullet\otimes_{k=1}^m p_k],
\end{equation}
for all $A\in \overline{{\cal A}}$, which includes distributional besides
smooth connections. $h_{e_i}(A)$ is an element of $SU(2)$ associated
with the edge $e_i$ and ``$\bullet$" stands for contracting, at each vertex $%
v_k$ of $\alpha$, the upper indices of the matrices corresponding to all the
incoming edges and the lower indices of the matrices assigned to the
outgoing edges, with all the indices of $p_k$.

Given a pair $\alpha,\vec\jmath \,$  the vector space generated by $T_{\alpha,%
{\vec \jmath},{\vec p}}$, for all possible contractors associated with $%
\alpha, \vec\jmath \, $ in the way stated previously,  is denoted by ${\cal H}^{\alpha,{\vec \jmath}%
}_{{\rm aux}}$. Then
\begin{equation}
{\cal H}_{{\rm aux}} = \oplus_{\alpha,{\vec \jmath}} {\cal H}^{\alpha,{%
\vec\jmath}}_{{\rm aux}},
\end{equation}
where $\alpha,{\vec \jmath}$ run over all the pairs consisting  of minimal graphs and
 irreducible non trivial representation labelings. The sum is orthogonal and the spaces
${\cal H}^{\alpha,{\vec\jmath}}_{{\rm aux}}$ are finite dimensional. It suffices to
define an orthonormal basis within each of them.

Note that the aforementioned construction of ${\cal H}_{{\rm aux}}$ holds actually
for any diffeomorphism covariant theory of connections with compact gauge
group. The choice of $SU(2)$ corresponds to  the case of  gravity described in terms of
real connection variables. So the generalization we are interested in to
include both gravity and the electromagnetic field is ${\cal H} = {\cal H}_{%
{\rm aux}} ^{SU(2)} \bigotimes {\cal H}_{{\rm aux}}^{ U(1)}$. The spin
network states for the compound system are denoted by $T_{\alpha,[{\vec
\jmath},{\vec p}],[\vec c, \vec q]} [A, \underline{A}]$, with $\vec c, \vec
q $ labeling the $U(1)$ coloring and contractors, respectively.

To extract physical information we will further need a state describing a flat
continuous space $\Sigma$ at scales much larger than the Planck length, but not
necessarily so at distances comparable to Planck length itself. States of this kind were introduced under the name of weave \cite{weave} for pure gravity.
 Flat weave
states $|W\rangle$, having a characteristic length ${\cal L}$, were first constructed
by considering collections of Planck scale circles randomly oriented. If one probes
distances $d >\!> {\cal L}$ the continuous flat classical geometry is regained,
while for distances $d <\!< {\cal L}$ the quantum loop structure of space is manifest.
In other words, one expects a behavior of the type $\langle W| {\hat q}_{ab}|W
\rangle= \delta_{ab} + O\left( \frac{\ell_P}{{\cal L}} \right)\,. $ It was soon
realized that such states could not yield a non trivial volume due to the lack of
self intersections \cite{intersecting}. Couples of circles, intersecting at a point,
were also considered as specific models of weaves to overcome this defect
\cite{twocircles}. With the recent advances on the kinematical Hilbert space ${\cal
H}_{{\rm aux}}$ it became clear that all proposed weaves were afflicted by two
undesirable features. First, they are defined to be peaked at a specific (flat or
curved) metric, but not at a connection. This is in
contrast with standard semiclassical states in terms of coherent
states, for example. Second, the known weave states do not belong either to $%
{\cal H}_{{\rm aux}}$ or to a dense subspace of it \cite{gaussweave}. It may be
possible to come to terms with such difficulties by  defining coherent
states for diffeomorphism covariant gauge theories \cite{aei} or by implementing a
genuine statistical geometry \cite{statg}, for instance.  Both approaches  have recently achieved
substantial progress.

 Nonetheless, in order to extract
physics, there is the alternative possibility of using just
the main features that { semiclassical states} should have . Namely, peakedness on both geometry and connection together
with the property that they yield well defined expectation values of physical
operators. An advantage of this alternative is that one may elucidate some physical
consequences before the full fledged semiclassical analysis is  settled down.
Indeed, such alternative may be considered as complementary, in the sense of hinting
at possible features of semiclassical states which could be further elaborated. After
its completion, a rigorous semiclassical treatment should tell us  whether the results
arising from this alternative turn out to hold or not. The weakness of the treatment
resides on its generality, since no detailed features of the { would be
semiclassical states} are used -as opposed, say, to the original weave states- and
hence a set of numerical coefficients cannot be calculated. Evaluating them will be
the task of the rigorous semiclassical treatment.

On top of the {purely gravitational semiclassical states}, a generalization
is required to include matter fields. For our analysis it will just suffice to
 exploit the same aspects of peakedness and well defined expectation
values, extended to include the case of the electromagnetic field. The { semiclassical states } here considered will describe flat space and a smooth electromagnetic field living in
it. Such a state is denoted by $|W,{\ \underline{\vec E}}, \underline{{\vec B%
}}>$ and has a characteristic length ${\cal L}$.  Since no detailed information is used on how the semiclassical
state is constructed in terms of, say, a graph, as opposed to weave states, the present
approach yields results relying only on the following assumptions: (i) peakedness of
the states, (ii) well defined expectation values and iii) existence of a
coarse-grained expansion involving ratios of the relevant scales of the problem: the
Planck length $\ell_P$, the characteristic length ${\cal L}$ and the electromagnetic
wavelength $\lambda$. States fulfilling  such requirements are referred to as {\it would be semiclassical states} in the sequel.

\section{The Regularization}

\label{thiereg}

Thiemann has put forward a consistent regularization procedure to define the
quantum Hamiltonian constraint of general relativity on ${\cal H}_{{\rm aux}%
} $, both for pure gravity and matter couplings \cite{Thiemann}. The basis of his
proposal is the incorporation of the volume operator as a convenient regulator, since
its action upon spin network states is finite. We use his regularization for the
Einstein-Maxwell theory, which naturally allows the semiclassical treatment here
pursued.

Consider the electric part of (\ref{hmax}). The identity $\frac{1}{\kappa}%
\left\{ A^i_a,V\right\}= 2{sgn}(\det e^j_b) e^i_a$ allows to rewrite it as
\begin{eqnarray}
H^{E}[N] &=& \frac{1}{2\kappa^2Q^2} \lim_{\epsilon\rightarrow 0} \frac{1}{%
\epsilon^3} \int_{\Sigma} d^3x N(x) \frac{\left\{A^i_a(x),V\right\}}{ 2(\det q)^{%
\frac{1}{4}} (x)} \underline{E}^a (x)  \nonumber \\
&&\mbox{}\times\, \int_{\Sigma} d^3y \chi_{\epsilon}(x,y) \frac{\left\{A^i_b(y),V%
\right\}}{ 2(\det q)^{\frac{1}{4}} (y)} \underline{E}^b (y),  \nonumber \\
&=& \frac{1}{2\kappa^2Q^2} \lim_{\epsilon\rightarrow 0} \int_{\Sigma} d^3x N(x)
\left\{A^i_a(x),\sqrt{V(x,\epsilon)})\right\} \underline{E}^a (x)  \nonumber
\\
&&\mbox{} \times \, \int_{\Sigma} d^3y \chi_{\epsilon}(x,y) \left\{A^i_b(y),\sqrt{%
V(y,\epsilon)}\right\} \underline{E}^b (y), \label{eq:HEreg}
\end{eqnarray}
with $\chi_{\epsilon}(x,y)= \Pi_{a=1}^{3} \theta (\epsilon/2-|x^a-y^a|)$ being
the characteristic function of a cube with volume $\epsilon ^3$ centered at $%
x$ and $V(x,\epsilon):= \int d^3y \chi_{\epsilon}(x,y)\sqrt{\det q}(y)$
being the volume of the box as determined by $q_{ab}$. Remarkably all $%
\epsilon$ dependence resides here. This is possible due to ${ H}_{{\rm %
Maxwell}}$ having density weight one and it is achieved at the price of explicitly
breaking diffeomorphism covariance. This is harmless as far as diffeomorphism
covariance is regained once the regulator is removed. This
is the case indeed \cite{Thiemann}. Next, let $\Sigma$ be triangulated into tetrahedra $%
\Delta$. Hence, the integral over $\Sigma$ in (\ref{eq:HEreg}) is just a sum over
the contributions of each  tetrahedron  $\Delta$.

The form of (\ref{eq:HEreg}) suggests to focus upon  the term inside each integral.
 As we will see below, this indeed simplifies the analysis. Let
\begin{eqnarray}
\Theta ^{i}[f] &:=&\int d^{3}x\;f(x){\underline{E}}^{a}(x)\left\{ A_{a}^{i}(x),%
\sqrt{V(x,\epsilon )}\right\}  \nonumber \\
&=&\sum_{\Delta }\int_{\Delta }d^{3}x\;f(x){\underline{E}}^{a}(x)\left\{
A_{a}^{i}(x),\sqrt{V(x,\epsilon )}\right\}  \nonumber \\
 &=&\sum_{\Delta }\int_{\Delta }f(x)e(x)\wedge \left\{ A^{i}(x),%
\sqrt{V(x,\epsilon )}\right\}  \nonumber \\
\Theta ^{i}[f]&=:& \sum_{\Delta }\Theta _{\Delta }^{i} [f]. \label{eq:ThetaDelta}
\end{eqnarray}%
Also let us use the dual of  $\underline E\,$:
$e_{bc}:= {^{\ast}{\underline E}}_{bc}=\frac{1}{1!}\epsilon _{bcd}{\underline{E}}%
^{d}$. For a two-surface $S$, $\int_{S}e
=\frac{1}{2}\int_{S}{\underline{E}}^{d}\;\epsilon _{bcd}\;dx^{b}\wedge dx^{c} =
\frac{1}{2}\int_{S}n_{d}{\underline{E}}^{d}\; \tilde{\epsilon}_{bc}\;dx^{b}\wedge
dx^{c} =\int_{S}{\underline{E}}^{d}n_{d}\;\tilde{\epsilon}$, $\tilde{\epsilon}$ being
the volume two-form. Hence $ \Phi ^{E}(S) :=\int_S e $ is the flux of
${\underline{E}}^{a}$ through $S$. Recalling that
\begin{eqnarray}
t{r}\left( \tau _{i}h_{s_{L}}\left\{ h_{s_{L}}^{-1},\sqrt{V(x,\epsilon )}%
\right\} \right) &=&t{r}\left( \tau _{i}\tau _{m}\int_{0}^{1}dt\;\dot{s}%
_{L}^{-1a}(t)\left\{ {A}_{a}^{m}(s_{L}^{-1}(t)),\sqrt{V(x,\epsilon )}%
\right\} \right) +\dots  \nonumber \\
&=&-\frac{\delta _{im}}{2}\int_{0}^{1}dt\;\dot{s}_{L}^{-1a}(t)\left\{ {A}%
_{a}^{m}(s_{L}^{-1}(t)),\sqrt{V(x,\epsilon )}\right\} +\dots  \nonumber \\
&\approx &-\frac{1}{2}s_{L}^{a}(1)\left\{ {A}_{a}^{i}(s_{L}^{-1}(0)),\sqrt{%
V(x,\epsilon )}\right\}
\end{eqnarray}
and that, for small tetrahedra, $ \Phi ^{E}(F_{JK}) \approx \frac{1}{2}\epsilon
_{abc}s_{J}^{b}(\Delta )s_{K}^{c}(\Delta ){\underline{E}}^{a}$, it follows
\begin{eqnarray}
&& f(v)\;\epsilon ^{JKL}\Phi ^{E}(F_{JK})\,tr\left( \tau ^{i}\;h_{s_{L}(\Delta
)}\left\{ h_{s_{L}(\Delta )}^{-1},\;\sqrt{V(v(\Delta
),\epsilon )}\right\} \right)  \nonumber \\
&\approx &-\frac{1}{4}f(v)\epsilon ^{JKL}\epsilon _{abc}s_{J}^{b}(\Delta
)s_{K}^{c}(\Delta ){\underline{E}}^{a}s_{L}^{d}(\Delta )\left\{ {A}%
_{d}^{i}(s_{l}^{-1}(0)),\sqrt{V(x,\epsilon )}\right\}  \nonumber \\
&=&-\frac{3!}{2}f(v)v{ol}(\Delta ){\underline{E}}^{a}\left\{ {A}%
_{a}^{i}(s_{l}^{-1}(0)),\sqrt{V(x,\epsilon )}\right\}  \nonumber \\
&=&-\frac{3!}{2}\int_{\Delta }f \, e\wedge \left\{ {A}^{i}(x),\sqrt{V(x,\epsilon )%
}\right\}\;.
\end{eqnarray}
We have then
\begin{equation}
\Theta _{\Delta }^{i}[f]=-\frac{2}{3!}\;f(v)\;\epsilon ^{JKL}\Phi
^{E}(F_{JK})tr\left( \tau ^{i}\;h_{s_{L}(\Delta )}\left\{ h_{s_{L}(\Delta
)}^{-1},\;\sqrt{V(v(\Delta ),\epsilon )}\right\} \right),  \label{eq:TE}
\end{equation}%
where we are denoting by\ $s_{J}(\Delta ),s_{K}(\Delta ),s_{L}(\Delta )$ the
edges of the tetrahedra\ $\Delta  $ having  $v$ as common vertex. As stated, $%
F_{JK}$ is a surface parallel to the face determined by $s_{J}(\Delta ),s_{K}(\Delta
)$ which is transverse to $s_{L}(\Delta )$.

Hence
\[
H^{E}[N]=\frac{1}{2\kappa ^{2}Q^{2}}\lim_{\epsilon \longrightarrow
0}\sum_{\Delta \Delta ^{\prime }}\Theta _{\Delta }^{i}[N]{\Theta ^{\prime }}%
_{\Delta ^{\prime }}^{i}[\chi].
\]

Next one replaces $\underline{E}^{a}$ and $V(x,\epsilon )$ by its quantum
counterparts and adapts the triangulation to the graph $\gamma$ corresponding to the state acted upon,
 in such a way that at each vertex $v$ of $\gamma$ and triplet of
edges $e,e^{\prime},e^{\prime\prime}$ a tetrahedron is defined with basepoint at the
vertex $v(\Delta)=v$ and segments $s_I(\Delta),\,
I=1,2,3$, corresponding to $s(e),s(e^{\prime}),s(e^{\prime%
\prime})\,$, respectively \cite{Thiemann}. Here it is assumed that $\epsilon^{IJK}\epsilon_{abc}{s_I}^a{s_J}^b{%
s_K}^c\geq 0$. The arcs connecting the end points of $s_I(\Delta)$ and $%
s_J(\Delta)$ are denoted $a_{IJ}(\Delta)$, so that a loop $\alpha_{IJ}:= s_I\circ
a_{IJ}\circ s_J^{-1}$ can be formed. Besides, the face spanned by the segments
$s_I(\Delta)$ and $s_J(\Delta)$ is called $F_{IJ}$.

The action of the regulated operator hereby obtained gets concentrated in the
vertices of the graph, as expected from the explicit appearance of the volume
operator. In successive steps we replace

\begin{equation}
\hat{\Theta}_{\Delta }^{i}[N]=-\frac{2}{3!}\frac{1}{i\hbar }\;N(v(\Delta))\;\epsilon
^{JKL}{\hat{\Phi}}^{E}(F_{JK})\,tr\left( \tau ^{i}\;h_{s_{L}(\Delta )}\left[
h_{s_{L}(\Delta )}^{-1},\;\sqrt{\hat{V}(v(\Delta ),\epsilon )}\right] \right)
\end{equation}%
\begin{equation}
\hat{\Theta}_{\Delta ^{\prime }}^{\prime i}[\chi]=-\frac{2}{3!}\frac{1}{i\hbar }%
\;\chi_{\epsilon}(v(\Delta),v(\Delta^{\prime}))\;\epsilon
^{MNP}{\hat{\Phi}}^{E}(F_{MN}^{\prime })\,tr\left( \tau
^{i}\;h_{s_{P}(\Delta ^{\prime })}\left[ h_{s_{P}(\Delta ^{\prime })}^{-1},\;%
\sqrt{\hat{V}(v(\Delta ^{\prime }),\epsilon )}\right] \right)
\end{equation}%
to obtain%
\begin{eqnarray}
H^{E}[N] &=&-\frac{1}{\hbar ^{2}2\kappa ^{2}Q^{2}}\sum_{v\in V(\gamma
)}\,N(v)\left( \frac{2}{3!}\frac{8}{E(v)}\right) ^{2}\sum_{v(\Delta )=v({%
\Delta ^{\prime }})=v}\times  \nonumber \\
&&\times \,tr\left( \tau ^{i}\;h_{s_{L}(\Delta )}\left[ h_{s_{L}(\Delta
)}^{-1},\;\sqrt{\hat{V}(v(\Delta ),\epsilon )}\right] \right) \,\epsilon
^{JKL}{\hat{\Phi}}^{E}(F_{JK})\times  \nonumber \\
&&\times \,tr\left( \tau ^{i}\;h_{s_{P}(\Delta ^{\prime })}\left[ h_{s_{P}(\Delta
^{\prime })}^{-1},\;\sqrt{\hat{V}(v(\Delta ^{\prime
}),\epsilon )}\right] \right) \,\epsilon ^{MNP}{\hat{\Phi}}%
^{E}(F_{MN}^{\prime }).\, \label{eq:HTT}
\end{eqnarray}
The valence $n(v)$ of the vertex $v$ yields the contribution $%
E(v)=n(v)(n(v)-1)(n(v)-2)/3!$ of the adapted triangulation at each vertex of $\gamma
$. Also, as $\epsilon \rightarrow 0$, $v(\Delta )=v(\Delta ^{\prime })$ are the only
contributions left over. The final expression for the electric piece of the Hamiltonian constraint  given in \cite{Thiemann} is obtained by the
explicit action of this operator on cylindrical functions. We refrain from doing that
here because the form of the operator (\ref{eq:HTT}) is better suited for our
approximation given below.

As for the magnetic part of $H_{\mathrm Maxwell}$ we proceed similarly. Since
\begin{eqnarray}
\underline{h}_{s} &=&{\rm e}^{-i\int_{0}^{1}dt\dot{s}^{a}(t)\underline{A}%
_{a}(s(t))}=\underline{I\!\!I}-i\int_{0}^{1}dt\dot{s}^{a}(t)\underline{A}%
_{a}(s(t))+\dots  \nonumber \\
(\underline{h}_{\alpha _{JK}}-1) &=&-i\int_{0}^{1}dt\;\dot{\alpha}%
_{JK}^{a}(t)\underline{A}_{a}(s(t))+\dots  \nonumber \\
&=&-i\int_{F_{JK}}\underline{B}^{a}dS_{a}+\dots  \nonumber \\
&=&-i\Phi ^{B}(F_{JK})+\dots  \nonumber \\
&\approx &-i\frac{1}{2}\epsilon _{abc}s_{J}^{\;b}(1)s_{K}^{\;c}(1)B^{a}(v(\Delta ))
\end{eqnarray}%
and
\begin{eqnarray}
&&f(v)\epsilon ^{JKL}(\underline{h}_{\alpha _{JK}}-1){tr}\left( \tau
_{i}h_{s_{L}(\Delta )}\left\{ h_{s_{L}(\Delta )}^{-1},\sqrt{V(x,\epsilon )}%
\right\} \right)  \nonumber \\
&\approx &i\frac{1}{4}\epsilon ^{JKL}\epsilon
_{abc}s_{J}^{\;b}s_{K}^{\;c}s_{L}^{\;d}f(v)B^{a}(v)\left\{ A_{d}^{i}(v),%
\sqrt{V(x,\epsilon )}\right\}  \nonumber \\
&=&i\frac{1}{2}vol(s_{J},s_{K},s_{L})\;\delta _{a}^{d}\;f(v)\;B^{a}(v)\left\{
A_{d}^{i}(v),\sqrt{V(x,\epsilon )}\right\}
\nonumber \\
&=&i\frac{3!}{2}vol(\Delta )f(v)B^{a}(v)\left\{ A_{a}^{i}(v),\sqrt{%
V(x,\epsilon )}\right\}  \nonumber \\
&=&i\frac{3!}{2}\int_{\Delta }f(x)\;B^{I}(x)\wedge \left\{ A^{i}(x),\sqrt{%
V(x,\epsilon )}\right\},
\end{eqnarray}
we can write
\begin{eqnarray} H^{B}[N] &=&\frac{1}{2\kappa ^{2}Q^{2}}\lim_{\epsilon
\longrightarrow 0}\sum_{\Delta \Delta ^{\prime }}\Xi _{\Delta }^{i}[N]\;\Xi _{\Delta
^{\prime}}^{\prime i}[\chi],
\end{eqnarray}
where
\begin{eqnarray}
\Xi _{\Delta }^{i}[f]&:=& i\frac{2}{3!}f(v)\epsilon ^{JKL}(\underline{h}%
_{\alpha _{JK}}-1){tr}\left( \tau _{i}h_{s_{L}(\Delta )}\left\{ h_{s_{L}(\Delta
)}^{-1},\sqrt{V(x,\epsilon )}\right\} \right).
\end{eqnarray}

Contrasting $\;\Xi _{I\Delta }^{i}$ with $\Theta _{I\Delta }^{i}$ we notice (i)
different numerical factors: $i\frac{2}{3!}$ for the former while $-\frac{2}{3!}$
for the latter and (ii) to leading order, the  magnetic flux $\epsilon ^{JKL}(\underline{h}%
_{\alpha _{JK}}-1)$  has as its counterpart  the
electric flux $\epsilon ^{JKL}E_{\alpha _{JK}}$.

The quantum version of the above operators is obtained using
\[
\hat{\Xi}_{\Delta }^{i}[f]:=i\frac{2}{3!}\frac{1}{i\hbar }\;f(v)\;\epsilon
^{JKL}(\underline{h}_{\alpha _{JK(\Delta )}}-1){tr}\left( \tau
_{i}h_{s_{L}(\Delta )}\left[ h_{s_{L}(\Delta )}^{-1},\sqrt{V(x,\epsilon )}%
\right] \right).
\]%
Finally, one gets the regularized magnetic piece of the Hamiltonian constraint as
\cite{Thiemann}
\begin{eqnarray}
H^{B}[N] &=&+\frac{1}{\hbar ^{2}2\kappa ^{2}Q^{2}}\sum_{v\in V(\gamma
)}\,N(v)\left( \frac{2}{3!}\frac{8}{E(v)}\right) ^{2}\sum_{v(\Delta )=v({%
\Delta ^{\prime }})=v}\times  \nonumber \\
&\times &\epsilon ^{JKL}\,tr\left( \tau _{i}\,h_{s_{L}(\Delta )}\left[
h_{s_{L}(\Delta )}^{-1},\sqrt{{\hat{V}}_{v}}\right] \right) \,\left( {%
\underline{h}}_{\alpha _{JK}(\Delta )}-1\right) \times  \nonumber \\
&\times &\epsilon ^{MNP}\,tr\left( \tau _{i}\,h_{s_{P}(\Delta ^{\prime })} \left[
h_{s_{P}(\Delta ^{\prime })}^{-1},\sqrt{{\hat{V}}_{v}}\right] \right) \,\left(
{\underline{h}}_{\alpha _{MN}(\Delta ^{\prime })}-1\right). \, \label{hm1}
\end{eqnarray}%
The electric and magnetic pieces of $H_{\mathrm Maxwell}$ can be treated in a unified
manner in terms of fluxes. To see this recall that for abelian gauge fields
\[
{\underline{h}}_{\alpha _{JK}(\Delta )}=e^{-i\int_{\alpha _{JK}(\Delta )}dt%
\dot{s}^{a}(t)\underline{\hat{A}}_{a}(s(t))}=e^{-i{\hat \Phi}^{B}(F_{JK})}\, ,
\]%
where ${\hat \Phi}^{B}(F_{JK})$ is the flux of the magnetic field through the surface
$F_{JK}$.

Then the full electromagnetic Hamiltonian is
\begin{eqnarray}
H_{\mathrm Maxwell}[N] &=&+\frac{1}{\hbar ^{2}2\kappa ^{2}Q^{2}}\sum_{v\in V(\gamma
)}\,N(v)\left( \frac{2}{3!}\frac{8}{E(v)}\right) ^{2}\sum_{v(\Delta )=v({%
\Delta ^{\prime }})=v}\times   \nonumber   \\
&&\times \,tr\left( \tau _{i}\,h_{s_{L}(\Delta )}\left[ h_{s_{L}(\Delta
)}^{-1},\sqrt{{\hat{V}}_{v}}\right] \right) tr\left( \tau
_{i}h_{s_{P}(\Delta ^{\prime })}\left[ h_{s_{P}(\Delta ^{\prime })}^{-1},%
\sqrt{{\hat{V}}_{v}}\right] \right) \,  \nonumber \\
&&\times \epsilon ^{JKL}\epsilon ^{MNP}\left[ \left( e^{-i{\hat{\Phi}}%
^{B}(F_{JK})}-1\right) \left( e^{-i{\hat{\Phi}}^{B}(F_{MN}^{\prime
})}-1\right) -\hat{\Phi}^{E}({F_{JK}})\hat{\Phi}^{E}({F_{MN}^{\prime }})%
\right].\nonumber \\ \label{EHC}
\end{eqnarray}
Let us emphasize the structure of the above regularized Hamiltonian. There is a common
gravitational factor included in the $SU(2)$ trace. The basic entities that
regularize the electromagnetic part are the corresponding fluxes: one is associated
with the magnetic field, which enters through a product of exponential flux factors, while
the other is related to the electric field, entering in a \ bilinear product of fluxes. Thus,
in the quadratic field approximation the effective Hamiltonian preserves duality
invariance. Nevertheless, the magnetic sector includes  higher powers in the field expansion. Hence, nonlinearities in the field equations, inducing
duality violations, arise only via the magnetic field.

Before proceeding let us recall that, according to
Thiemann's conventions,
in flat space we must have
\begin{equation}
H_{\mathrm Maxwell}=\int d^{3}x\frac{1}{2\,Q^{2}}\left( \underline{E}^{a}\underline{E}^{a}+%
\underline{B}^{a}\,\underline{B}^{a}\right) ,\quad  \label{CLASLIM}
\end{equation}%
where $Q$ is the electromagnetic coupling constant. The electromagnetic potential is
denoted by ${\underline{A}}_{a}\,\, $and the corresponding electromagnetic tensor by
$\underline{F}_{ab}$. The units are such that the gravitational connection
$A_{a}^{i}$ has dimensions of $1/L$ (inverse Length) and the
Newton's constant $\kappa $ has dimensions of $L/M$ (Length over Mass). Also we have that $[|%
\underline{{\vec{E}}}|/Q^{2}]=M/L^{3}$. Taking the dimensions of ${%
\underline{A}}_{a}$ to be $1/L$, according to the corresponding
normalization  of the Wilson loop, we conclude that  $[\underline{{%
\vec{E}}}]=[\underline{{\vec{B}}}]=1/L^{2}$ and   $%
[Q^{2}]=1/(M\,L)$. In our case we also have $[\hbar ]=M\,L$, which in fact leads to
$\alpha _{EM}=Q^{2}\,\hbar $ to be the dimensionless fine-structure constant, as
defined by Thiemann \cite{Thiemann}.

\section{ General structure of the calculation}

\label{GIC}

The effective Maxwell Hamiltonian is defined by considering the expectation value of
the $U(1)$ gauge sector of the quantum Hamiltonian constraint with respect to
$|W,\underline{{\vec E}}, \underline{{\vec B}}\rangle$. Inside this expectation
value  operators are expanded around all relevant vertices of the triangulation in
powers of the segments $s^a_L(\Delta)$, which have lengths of  order $\ell_P$. In this
way, a systematic approximation is giv\-en involving the scales $\ell_P <\!\!< {\cal
L}\, < \,\lambda$, where  $\lambda$ is the De Broglie wavelength of the photon. Our corrections to the Maxwell Hamiltonian arise from such an approximation.

We do the full calculation of the magnetic sector, including  the non-linear contributions  to order $\ell_P^2$. Next, to obtain the electric sector, it is enough to consider only the
quadratic terms in the magnetic Hamiltonian and make the replacement $B \rightarrow E.$

In the case of the magnetic  sector, the general form of the expectation value is
(recalling that $1/\kappa ^{2}=\hbar ^{2}/\ell _{P}^{4}$ )
\begin{eqnarray}
{\rm H}^{B}=-\frac{1}{2\,Q^{2}}\frac{1}{\ell _{P}^{4}}\,\sum_{v\in V(\gamma
)}N(v)\left( \frac{2}{3!}\frac{8}{\,E(v)}\right) ^{2}\sum_{v(\Delta
)=v({\Delta ^{\prime }})=v}\,<W,\,\underline{{\vec{E}}},\,\underline{{\vec{B}%
}}|\underline{{\hat{F}}}_{p_{1}\,q_{1}}(v)\dots \,\underline{{\hat{F}}}%
_{p_{n}\,q_{n}}(v)\, &&  \nonumber  \label{GENFM} \\
\,\left( \partial ^{a_{1}}\dots \partial ^{a_{m}}\,\underline{{\hat{F}}}%
_{pq}(v)\right) \,\,{\hat{T}}_{a_{1}\dots a_{m}}{}^{\,pq\,p_{1}\,q_{1}\dots
p_{n}\,q_{n}}(v,s(\Delta ),\,s(\Delta ^{\prime }))|W,\,\underline{{\vec{E}}}%
,\,\underline{{\vec{B}}}>. &&
\end{eqnarray}

To proceed with the approximation we think of space as made up of boxes, each centered
at a given  point ${\vec{x}}$ and with volume ${\cal L}^{3}\approx d^{3}\,x$. Each box
contains a large number of vertices of the semiclassical state (${\cal L}>\!>\ell
_{P}$), but is considered as infinitesimal in the scale where the space can be regarded
as continuous. Also, we assume that the magnetic operators are slowly varying inside
the box ($\ell _{P}<\!<\lambda _{D}$), in such a way that for all the vertices inside
the box one can write
\begin{equation}
\langle W,\,\underline{{\vec{E}}},\,\underline{{\vec{B}}}|\dots \underline{{%
\hat{F}}}_{ab}(v)\dots |W,\,\underline{{\vec{E}}},\,\underline{{\vec{B}}}%
\rangle = \mu \,\epsilon_{abc}\underline{B}^{c}({\vec{x}}).
\end{equation}%
Here ${\ {\underline F}}_{ab}=\partial_a\,{\underline A}_b-\partial_b\,{%
\underline A}_a$,\, $\underline{B}^a({\vec x})$ is the classical magnetic field at
the center of the box and $\mu$ is a dimensionless constant to be
determined in such a way that we recover the standard classical result (\ref%
{CLASLIM}) in the zeroth order approximation.
 In the next section we show that
\begin{equation}
\mu =\left(\frac{{\cal L}}{\ell_P}\right)^{1+ \Upsilon}\;,
\end{equation}
with $\Upsilon$ being a parameter defining  the leading order contribution of the
gravitational connection to the expectation value.
 Applying to (\ref{GENFM}) the procedure just described leads to
\begin{eqnarray}
{\rm H}^{B}&=&\sum_{{\rm Box}}\,N({\vec{x}})\,\underline{{F}}_{p_{1}\,q_{1}}({\vec{x}}%
)\dots \,\underline{{\ F}}_{p_{n}\,q_{n}}({\vec{x}})\dots \,\left( \partial
^{a_{1}}\dots \partial ^{a_{m}}\,\,\underline{{\ F}}_{p\,q}({\vec{x}}%
)\right) \,\,\sum_{v\in {\rm Box}}\ell _{P}^{3}\,\left(\frac{2}{3!} \frac{8}{%
E(v)}\right) ^{2}  \nonumber \\
&& \sum_{{v(\Delta )=v({\Delta ^{\prime }})=v}}\,\left( -\frac{1}{2\,Q^{2}}\frac{\mu ^{n+1}}{\,\ell _{P}^{4}}\right) <W,\,%
\underline{{\vec{E}}},\,\underline{{\vec{B}}}|\frac{1}{\ell _{P}^{3}}{\hat{T}%
}_{a_{1}\dots a_{m}}{}^{\,pqp_{1}\,q_{1}\dots p_{n}\,q_{n}}(v,s(\Delta ),\,s(\Delta
^{\prime }))|W,\,\underline{{\vec{E}}},\,\underline{{\vec{B}}}>
\nonumber \\
&=&\sum_{{\rm Box}}\ N({\vec{x}})\,\underline{{\ F}}_{p_{1}\,q_{1}}({\vec{x}}%
)\dots \,\underline{{\ F}}_{p_{n}\,q_{n}}({\vec{x}})\left( \frac{{}}{{}}%
\partial ^{a_{1}}\dots \partial ^{a_{m}}\underline{{\ F}}_{pq}({\vec{x}})%
\frac{{}}{{}}\right) \,\ d^{3}x\ {{T}}_{a_{1}\dots
a_{m}}{}^{\,pqp_{1}\,q_{1}\dots p_{n}\,q_{n}}({\vec{x}})  \nonumber \\
{\rm H}^{B} &=&\int d^{3}x\ N({\vec{x}})\underline{{\ F}}_{p_{1}\,q_{1}}({%
\vec{x}})\dots \,\underline{{\ F}}_{p_{n}\,q_{n}}({\vec{x}})\,\left( \frac{{}%
}{{}}\partial ^{a_{1}}\dots \partial ^{a_{m}}\underline{{\ F}}_{pq}({\vec{x}}%
)\right) \,\,{{T}}_{a_{1}\dots a_{m}}{}^{\,pqp_{1}\,q_{1}\dots
p_{n}\,q_{n}}({\vec{x}}).  \label{AMH}
\end{eqnarray}%
The box-averaged tensor ${{T}}_{a_{1}\dots a_{m}}{}^{\,pqp_{1}\,q_{1}\dots
p_{n}}({\vec{x}})$, defined by
\begin{eqnarray}
{{T}}_{a_{1}\dots a_{m}}{}^{\,pqp_{1}\,q_{1}\dots p_{n}\,q_{n}}({\vec{x}}%
)=\sum_{v\in {\rm Box}}\,\left( \frac{2}{3!}\frac{8}{E(v)}\right)
^{2}\sum_{v(\Delta )=v({\Delta ^{\prime }})=v}\,\,\left(- \frac{1}{2\,Q^{2}}\frac{\mu ^{n+1}}{%
\,\ell _{P}^{4}}\right) &&  \nonumber  \label{BAMT} \\
\times \,<W,\,\underline{{\vec{E}}},\,\underline{{\vec{B}}}|\frac{1}{\ell
_{P}^{3}}{\hat{T}}_{a_{1}\dots a_{m}}{}^{\,pqp_{1}\,q_{1}\dots
p_{n}\,q_{n}}(v,s(\Delta ),\,s(\Delta ^{\prime }))|W,\,\underline{{\vec{E}}}%
,\,\underline{{\vec{B}}}>, &&
\end{eqnarray}%
is constructed from flat space tensors like $\delta _{ab},\,\epsilon _{abc}$. In this
way we are demanding covariance under rotations at the scale ${\cal L}$.

When averaging inside each box, the scaling of the
expectation values of the gravitational operators is estimated according to
\begin{equation}
\langle W,\,\underline{{\vec{E}}},\,\underline{{\vec{B}}}\,|\dots
A_{ia}\,\dots  |W,\,\underline{{\vec{E}}},\,\underline{{%
\vec{B}}}\,\rangle \approx \dots \frac{1}{{\cal L}}\left( \frac{\ell _{P}}{%
{\cal L}}\right) ^{\Upsilon }\,\dots , \label{ESTGRAVOP}
\end{equation}
\begin{equation}
\langle W,\,\underline{{\vec{E}}},\,\underline{{\vec{B}}}\,|\dots
\sqrt{V_{v}} \,\dots  |W,\,\underline{{\vec{E}}},\,\underline{{%
\vec{B}}}\,\rangle \approx \dots \,\ell _{P}^{3/2}\dots \,, \label{ESTGRAVOP1}
\end{equation}
respectively. In our previous work \cite{URRU1} we have
set  $\Upsilon =0$ on the basis that the coarse graining approximation, defined by the scale ${\cal L}$, does not allow
for the connection to be probed below $\frac{1}{\cal L}$. On the other hand, by
adopting naive  kinematical coherent states for representing semiclassical states, one
would set $\Upsilon =1$ for two reasons: (i) to guarantee that (\ref{ESTGRAVOP}) yields just zero in the limit $\hbar\rightarrow 0$
, in agreement with a flat connection and (ii)
because such an scaling would saturate the Heisenberg uncertainty relation \cite{HEITLER}. Nonetheless,
physical semiclassical states may imply a leading order contribution with $\Upsilon\neq
0,1$, thus we choose to consider $\Upsilon$ as a free parameter here.

In order to make the transition to the electric sector it is convenient to express
the effective Hamiltonian ${\rm H}^B$ in terms of
the magnetic field, which amounts to a redefinition of the expression (\ref%
{AMH}) in the form
\begin{eqnarray}  \label{AMH1}
{\rm H}^B=\int d^3 x \ N({\vec x})\, \underline{{B}}_{r_1}({\vec x})\dots \,%
\underline{{B}}_{r_n}({\vec x}) \, \left( \frac{}{} \partial^{a_1}\dots
\partial^{a_m} \underline{{\ B}}_{r}({\vec x})\right)\,\, {R}_{a_1\dots
a_m}{}^{ \, r r_1 \dots r_n}({\vec x}).
\end{eqnarray}
The relation between the  box-averaged tensors $R$ and $T$ is
\begin{eqnarray}
{T}_{a_1\dots a_m}{}^{ pqp_1\,q_1\dots p_n q_n}({\vec x})= \epsilon^{pq}{}_r\,
\epsilon^{p_1q_1}{}_{r_1}\,\dots
\epsilon^{p_nq_n}{}_{r_n}\, {R}_{a_1\dots a_m}{}^{ \, r r_1 \dots r_n}({%
\vec x}).
\end{eqnarray}

By expanding (\ref{EHC}) to different  powers in $s_I^a(\Delta)$ one
can systematically determine all possible contributions to the effective electromagnetic Hamiltonian at a given order in $%
\ell_P$.

\section{The Calculation}

\label{thieregM}

In this section we provide the details of the calculation of the Maxwell effective Hamiltonian up to order $\ell_P^2$. Let us start with the magnetic sector.

The two main ingredients in (\ref{EHC}) which contribute to the expansion in powers of
the segments $s_I(\Delta)$ are: (i) the  trace factors involving the
gravitational operators and (ii) the magnetic flux through  each surface
$F_{IJ}(\Delta)$.

First we calculate the flux of the magnetic field through the  surface $%
F_{IJ}$. A convenient way to do this is via the Stokes theorem%
\begin{eqnarray}
\Phi ^{B}(F_{IJ}) &=&\int_{F_{IJ}}{B}_{a}n^{a}\;d^{2}x=\int_{F_{IJ}}\left(
\nabla \times \vec{A}\right) _{a}n^{a}\;d^{2}x \nonumber \\
&=&\int_{\alpha _{IJ}}dt\;\dot{s}^{a}(t)A_{a}(t) \nonumber \\
&=&\int_{{\vec{v}}}^{{\vec{v}}+{\vec{s}}_{I}}{\underline{A}}%
_{a}\,dx^{a}+\,\int_{{\vec{v}}+{\vec{s}}_{I}}^{{\vec{v}}+{\vec{s}}_{J}}{%
\underline{A}}_{a}\,dx^{a}+\int_{{\vec{v}}+{\vec{s}}_{J}}^{{\vec{v}}}{%
\underline{A}}_{a}\,dx^{a}. \label{STOKES}
\end{eqnarray}
Here the notation is ${\vec{s}}_{I}=\{s_{I}^{a}\}$ and analogously for ${%
\vec{v}}$. We are using  straight line trajectories joining the vertices of the
corresponding triangle.

The basic building block in (\ref{STOKES}) is
\begin{eqnarray}
\int_{{\vec{v}}_{1}}^{{\vec{v}}_{2}}{\underline{A}}_{a}({\vec{x}})\,dx^{a}
&=&\int_{0}^{1}{\underline{A}}_{a}({\vec{v}}_{1}+t\,({\vec{v}}_{2}-{\vec{v}}%
_{1}))\,({\vec{v}}_{2}-{\vec{v}}_{1})^{a}\,dt  \nonumber \\
&=&\int_{0}^{1}{\underline{A}}_{a}({\vec{v}}_{1}+t\,\vec{\Delta})\,\Delta
^{a}\,dt  \nonumber \\
&=&\left( 1+\frac{1}{2!}\Delta ^{b}\partial _{b}+\frac{1}{3!}(\Delta ^{b}\partial
_{b})^{2}+\dots \right) \Delta ^{a}{A}_{a}(v),
\end{eqnarray}%
with$\;\Delta ^{a}=({\vec{v}}_{2}-%
{\vec{v}}_{1})^{a}$. The infinite series in parenthesis is
\begin{equation}
F(x)=1+\frac{1}{2!}x+\frac{1}{3!}x^{2}+\frac{1}{4!}x^{3}+\dots =\frac{e^{x}-1%
}{x},  \label{FF}
\end{equation}%
yielding
\be
\int_{{\vec{v}}_{1}}^{{\vec{v}}_{2}}{\underline{A}}_{a}({\vec{x}}%
)\,dx^{a}=F(\Delta ^{a}\,\partial _{a})\left( \Delta ^{a}\,{\underline{A}}%
_{a}({\vec{v}}_{1})\right) .\quad \ee
In the following we employ the notation $\Delta
^{a}\,V_{a}={\vec{\Delta}}\cdot {\vec{V}}$. Using the above result in the three
integrals appearing in (\ref{STOKES}) and after some algebra, we obtain
\begin{eqnarray}
\Phi ^{B}(F_{IJ}) &=&F_{1}({\vec{s}}_{I}\cdot {\nabla},{\vec{s}}%
_{J}\cdot {\nabla})\,\,s_{J}^{a}\,s_{I}^{b}\,\left( \partial _{a}\,{%
\underline{A}}_{b}({\vec{v}})-\partial _{b}\,{\underline{A}}_{a}({\vec{v}}%
)\right) \nonumber \\
&=&F_{1}({\vec{s}}_{I}\cdot {\nabla},{\vec{s}}_{J}\cdot {\nabla}%
)\,\,s_{J}^{a}\,s_{I}^{b}\,\epsilon _{abc}B^{c}(v),
\end{eqnarray}%
where the gradient acts upon the coordinates of ${\vec{v}}$. The function $%
F_{1}$ is \be
F_{1}(x,y)=\frac{y({\rm e}^{x}-1)-x({\rm e}^{y}-1)}{x\,y\,(y-x)}%
=-\sum_{n=1}^{\infty }\frac{1}{(n+1)!}\,\frac{x^{n}-y^{n}}{x-y}. \ee
Let us emphasize that $F_{1}(x,y)$ is just a power series in the variables $x$ and $y$.
Expanding to fourth order in the segments $s_{I}^{a}$ we obtain
\begin{eqnarray}
\Phi ^{B}(F_{IJ}) &=&\left( 1\,\,+\frac{1}{3}(s_{I}^{c}+s_{J}^{c})\,\partial
_{c}\,+\frac{1}{12}(s_{I}^{c}\,s_{I}^{d}+s_{I}^{c}\,s_{J}^{d}+s_{J}^{c}%
\,s_{J}^{d})\,\,\partial _{c}\,\partial _{d}+...\right) \times  \nonumber
\label{EXPX} \\
&&\times \,\frac{1}{2}s_{I}^{a}s_{J}^{b}\epsilon _{abc}B^{c}(v).
\label{FLUX}
\end{eqnarray}%
Notice that the combination%
\be \frac{1}{2}s_{I}^{a}s_{J}^{b}\epsilon _{abc}={\cal A}\;n_{c} \ee
is just the oriented area of the triangle with vertex $v$ and sides $%
s_{I}^{c},\;s_{J}^{c}$, joining at this vertex, having value ${\cal A}$ and unit
normal$\;$vector$\;n_{c}$.

In order to make the bookkeeping clear, let us denote by $T$ the  combination
arising from  (\ref{EHC})
whose expectation value we are calculating
\ba
\label{T}
T=-\frac{1}{2\,Q^2\,
}\frac{1}{\ell_P^4}\, {\hat w}_{i \, L \Delta }{\hat
w}_{i \, P \Delta' }\, \epsilon^{JKL}\,\epsilon^{MNP}  \left( e^{-i{\hat{\Phi}}%
^{B}(F_{JK})}-1\right) \left( e^{-i{\hat{\Phi}}^{B}(F_{MN}^{\prime })}-1\right). \ea
 Here
\begin{eqnarray}
{\hat w}_{i \, L \Delta } &=& tr \left(\tau_i \, h_{s_L(\Delta)}
\left[h_{s_L(\Delta)}^{-1}, \sqrt{\hat{V}_v}  \right] \right). \label{WILD}
\end{eqnarray}

Some remarks are in order before we proceed further.
Our final goal is
to obtain a power expansion of ({\ref T}) up to order $\ell_P^2$. Since, as we will
show in the sequel, the normalization factor converting magnetic operators inside the
semiclassical expectation value into classical fields outside the expectation value is proportional to
$(\ell_P)^{-1}$, we have to take some care regarding the expansion of the given
quantities in powers of $s_I^a$. A detailed power counting analysis in the expression
({\ref T}) shows that the term ${\hat w}_{i\,L(\Delta)}$ in (\ref{WILD}) is to be
expanded up to order $s^3$, while each magnetic factor
$\left( e^{-i{\hat{\Phi}}%
^{B}(F_{JK})}-1\right)$ is required to have the following properties: the terms
proportional to $\underline{F}$ are to be expanded up to order $s^4$, those
proportional to $\underline{F}^2$ up to order $s^5$ and finally those proportional to
$\underline{F}^3$ up to order $s^6$. This will lead to the following contributions in
$T$: the terms proportional to $\underline{F}^2$ include the expansion up to order
$s^8$,the terms proportional to $\underline{F}^3$ include the expansion up to order
$s^9$ and the terms proportional to $\underline{F}^4$ include the expansion up to
order $s^{10}$. The final result is that the semiclassical expectation value of the
magnetic contribution $T$ will be proportional to $\ell_P^3$, which is incorporated
in the volume element, times corrections up to order $\ell_P^2$.

Let us now continue with the calculation of the   contribution  to (\ref{T})
due to the magnetic flux by writing the expansion
 \ba \label{WLC} \left(
e^{-i{\hat{\Phi}}^{B}(F_{JK}(\Delta))}-1\right) &=& \sum_{n=1}^{\infty}
\frac{(-i)^n}{n!} ({\hat{\Phi}}^{B}(F_{JK}))^n= M_{1JK(\Delta)} + M_{2JK(\Delta)} +
M_{3JK(\Delta)} \nonumber \\&&+ {\cal O}\left(s^7\underline{F}^3\right) \ea where
\ba \label{MJKD}
M_{1JK(\Delta)}&:=& s^a_K s^b_J \frac{i}{2!}  \underline{F}_{ab}, \nonumber\\
M_{2JK(\Delta)}&:=& s^a_K s^b_J \frac{i}{3!} (x_J+x_K)  \underline{F}_{ab}
           -s^a_K s^b_Js^c_K s^d_J \frac{1}{8}  \underline{F}_{ab}                                 \underline{F}_{cd},\nonumber \\
M_{3JK(\Delta)}&:=& s^a_K s^b_J \frac{i}{4!} (x_J^2+x_Jx_K+x_K^2)  \underline{F}_{ab}
-s^a_K s^b_Js^c_K s^d_J\left[\frac{1}{4\cdot 3!} (x_J+x_K)\underline{F}_{ab}\underline{F}_{cd} \right.\nonumber\\
&&\mbox{} \left. + \frac{1}{4\cdot 3!}
\underline{F}_{ab}(x_K+x_J)\underline{F}_{cd}\right] - s^a_K s^b_J s^c_K s^d_J s^e_K
s^f_J \frac{i}{4\cdot 3!} \underline{F}_{ab} \underline{F}_{cd} \underline{F}_{ef},
\ea according to the previous analysis. We are using the notation $x_I={\vec
s}_I\cdot\nabla=s_I^a\,\partial_a$. Let us remark that, contrary to the electric
case, the magnetic contribution will incorporate non-linear terms due to the
expansion of the exponential in powers of ${\vec B}$. This implies that the exact duality
symmetry of Maxwell equations in vacuum will be lost due to quantum gravity
corrections.

Next let us consider the gravitational contributions to (\ref{EHC}), arising from
({\ref {WILD}), which we expand as
\ba \label{WILD1} {\hat w}_{i \, L \Delta }&=&
s^a_L w_{ia} + s^a_Ls^b_L w_{iab} + s^a_Ls^b_Ls^c_L w_{iabc} + {\cal O}(s^4w), \ea with
\ba w_{ia} = \frac{1}{2} [A_{ia},\sqrt{V}],\quad w_{iab} = \frac{1}{8} \epsilon_{ijk}
[A_{ja},[A_{kb},\sqrt{V}]], \quad w_{iabc} = -\frac{1}{48}
[A_{ja},[A_{jb},[A_{ic},\sqrt{V}]]]. \ea
The scaling properties of the above gravitational operators under the
semiclassical expectation value are
 \be \langle
W\,\underline{{\vec E}}\,\underline{{\vec B}}| \dots w_{i \,a_1 \dots a_n} \dots
|W\,\underline{{\vec E}}\,\underline{{\vec B}} \rangle \rightarrow
\frac{\ell_P{}^{3/2}}{{\cal L}^n} \left( \frac{\ell_P}{{\cal L}}\right)^{n \Upsilon}.
\label{eq:Ascale} \ee
Let us emphasize that the result (\ref{eq:Ascale}) is a consequence of the  scaling of the expectation value  of the
connection given by (\ref{ESTGRAVOP}).

For the product ${\hat w}_{i \, L \Delta }{\hat w}_{i \, P \Delta' }$ we need only
\ba {\hat w}_{i \, L \Delta }{\hat w}_{i \, P \Delta' }&=& U_{2LP} + U_{3LP} +
U_{4LP} + {\cal O}\left(s^5w^2\right) \ea with \ba
U_{2LP} &=& s^a_L s'^d_P w_{ia}w_{id}, \nonumber \\
U_{3LP} &=& s^a_Ls'^d_Ps'^e_Pw_{ia}w_{ide} + s^a_Ls^b_Ls'^d_P w_{iab} w_{id}, \nonumber \\
U_{4LP} &=& {s^a_L} {s'^d_P} {s'^e_P} {s'^f_P} w_{ia} w_{idef}+ s^a_L s^b_L s'^d_P
s'^e_P w_{iab} w_{ide} + s^a_L s^b_L s^c_L s'^d_P w_{iabc} w_{id}. \label{w2} \ea
Here all the $w$'s are evaluated at a common vertex $v$.

At this level it is convenient to state the result (no sum over $L$) \ba
s^a_L \, s^b_L\, w_{iab}&=&\frac{1}{8}\, s^a_L\,s^b_L\, \epsilon_{ijk}\, [A_{ja},\, [A_{kb}, \, \sqrt{V}]]\nonumber \\
&=&\frac{1}{8}\, s^a_L\,s^b_L\, \epsilon_{ijk}\, \left( A_{ja}A_{kb}\sqrt{V}-
A_{ja}\sqrt{V}A_{kb}- A_{kb}\sqrt{V}A_{ja}+
\sqrt{V}A_{kb}A_{ja}\right), \nonumber \\
&=&0, \ea which holds due to symmetry properties. This leads to \be U_{3LP}=0. \ee

After taking the  expectation
value the terms contributing to order  $\ell_P^2$ in  (\ref{T}) read   \ba
T&=&T_0 + T_1 + T_2 + {\cal O}(\rightarrow \ell_P^3)\\
T_0 &=& - \frac{1}{2\,Q^2\, }\frac{1}{\ell_P^4}\, \epsilon^{JKL} \,\epsilon^{MNP}
\left[ U_{2LP}\, M_{1JK}\, M'_{1MN} \frac{}{}\right],\label{R0} \\
T_1 &=&-\frac{1}{2\,Q^2\,}\frac{1}{\ell_P^4}\, \epsilon^{JKL}\, \epsilon^{MNP}
 U_{2LP}\, \left[ \frac{}{} M_{1JK} \, M'_{2MN} + M_{2JK}\,  M'_{1MN}\right], \label{R1}\\
T_2 &=& - \frac{1}{2\,Q^2\,}\frac{1}{\ell_P^4}\, \epsilon^{JKL}\, \epsilon^{MNP}
\left[\frac{}{} U_{2LP}\, (M_{1JK}\, M'_{3MN}+ M_{3JK}\, M'_{1MN}  + M_{2JK}\, M'_{2MN})\,  + \right.\nonumber \\
&&\qquad \qquad \qquad \qquad \qquad \qquad \left.  + U_{4LP}\, M_{1JK}\,
M'_{1MN}\frac{}{} \right].\label{R2} \ea

Now we are ready to calculate the different contributions to the  magnetic sector of
the Hamiltonian (\ref{EHC}), which we parameterize in terms of the tensor ${\bar
W}_{a_1\dots a_m}{}^{r r_1\dots r_n}$ introduced in (\ref{BAMT}).

Recalling that  we are only interested in  the pieces which are symmetric in the
indices $r_1, r_2, \dots, r_n$, the contribution $T_0$ produces
\ba R_0{}^{
r_1r_2} &=& \sum_{v \in \, {\rm Box}}\, \frac{1}{2 Q^2}\,
\left(\frac{2}{3!}\frac{8}{E(v)}\right)^2 \, \sum_{v(\Delta)=v({\Delta'})=v}
\,\frac{\mu^2}{4\,\ell_P^7}\,\,\epsilon^{r_1}{}_{ab} \, \epsilon^{r_2}{}_{uv}
\, \epsilon^{JKL}\, \epsilon^{MNP} \times \nonumber \\
&& \times \, \, s_K^a\, s_J^b \,s_L^c\,\, {s'}_M^v \, {s'}_N^u\, {s'}_P^d  \ \langle
W\,\underline{{\vec E}}\,\underline{{\vec B}}|w_{ic}\,w_{id}\, |W\,\underline{{\vec
E}}\,\underline{{\vec B}}\rangle.
\label{W0R1R2}
\ea
In order to simplify the
product  of vectors $s_L^a$ (${s'}_M{}^p$) appearing in the sequel and also to
exhibit the internal symmetry properties of the quantities involved, it is  convenient to keep in mind
the relations
\be
\label{DET}
\epsilon^{KJL}\, s_K^a\, s_J^b \,s_L^c = det(s)\,
\epsilon^{abc}, \quad det(s)=det(s^a_K), \qquad \epsilon^{abp}\epsilon_{abq}=2\delta^p_q.
\ee
In this way,  (\ref{W0R1R2}) can be
rewriten in the simpler form
\ba R_0{}^{ r_1r_2} &=& \sum_{v \in \, {\rm
Box}}\, \frac{1}{2\,Q^2 }\, \left(\frac{2}{3!}\frac{8}{E(v)}\right)^2 \,
\sum_{v(\Delta)=v({\Delta'})=v} \,\frac{\mu^2}{\ell_P^7}\,
\,  \times \nonumber \\
&& \times \, \,
 det(s)\,\,
 det(s')\, \ \langle W,\,\underline{{\vec E}},\,\underline{{\vec B}}|\frac{1}{2}\{w_{i}{}^{r_1}, \,w_{i}{}^{r_2}\}\, |W,\,\underline{{\vec E}},\,\underline{{\vec B}}\rangle.
\ea The above equation implies \ba R_0{}^{ r_1r_2} &=& \frac{1}{2\,Q^2}\frac{\mu^2}{
\ell_P^7}\, \, \ell_P^6\, \frac{\ell_P^3}{{\cal L}^2} \left(\frac{\ell_P}{{\cal
L}}\right)^{2\Upsilon}\,\, \delta^{r_1r_2}=\frac{1}{2\,Q^2}\delta^{r_1r_2}, \ea which
reproduces the zeroth-order  magnetic contribution (\ref{CLASLIM}) with the choice \be
\mu= \left(\frac{{\cal L}}{\ell_P}\right)^{1+\Upsilon}. \ee

Now let us consider the correction arising from $T_1$, which leads to the following
contribution in the effective Hamiltonian
\ba &&{\rm H}^B_{11}=\sum_{{\rm Box}({\vec
x})}   \sum_{v \in \, {\rm Box}}\,  \frac{1}{2 Q^2\,} \left(\frac{2}{3!}
\frac{8}{E(v)}\right)^2 \,\ell_P^3 \sum_{v(\Delta)=v({\Delta'})=v}
-\,\frac{1}{\ell_P^7}\,  \epsilon^{JKL}\, \epsilon^{MNP}\,s^c_L s'^d_P \,  s^a_K
s^b_J
\times \nonumber \\
&& \times \,\langle W,\,\underline{{\vec  E}},\, \underline{{\vec  B}}|
\{w_{ic}\,,\,w_{id}\} \, \frac{i}{2}\, \underline{F}_{ab} \, \left(\frac{}{}{s'}^v_N
{s'}^q_M \frac{i}{3!} ({x'}_M+{x'}_N) \underline{F}_{vq}-\right. \nonumber \\
&& \left. \qquad \qquad \qquad \qquad \qquad \qquad
\qquad \qquad \qquad-{s'}^v_N {s'}^q_M {s'}^r_N
{s'}^t_M \frac{1}{8} \underline{F}_{vq}
\underline{F}_{rt}\right) \, |W, \underline{{\vec  E}},\, \underline{{\vec  B}}\rangle.
\ea
In the above expression we have interchanged the summations over
$\Delta$ and $\Delta'$ in order to rewrite $w_{ic}\,w_{id}$ as
$\frac{1}{2}\{w_{ic}\,,\,w_{id}\} $. We further separate the above contribution to
${\rm H}^B_{11}$ in two parts: (i) the first contains two powers in the magnetic
field and leads to $R_{11a_1}{}^{rr_1}$ and (ii) the second one contains
three powers in the magnetic field and leads to the completely symmetric tensor
$R_{11}{}^{r_1r_2r_3}$ . Since we have no  symmetric tensor with three indices
at our disposal, the latter contribution is zero. Thus we concentrate in the first one
\ba
R_{11}{}^{a_1rr_1}&=& \sum_{v \in \, {\rm Box}}\,  \frac{1}{2\,Q^2}\, \left(\frac{2}{3!}\frac{8}{E(v)}\right)^2 \, \sum_{v(\Delta)=v({\Delta'})=v}\frac{\mu^2}{6\,\ell_P^7}\, \epsilon^{JKL}\, \epsilon^{MNP}\,s^c_L s'^d_P \,\times \nonumber \\
&& \, s^a_K s^b_J \, {s'}^v_N {s'}^q_M \, {s'}_M^{a_1}\, \epsilon^{r_1}{}_{ab}\,
\epsilon^r{}_{vq} \langle W,\, \underline{{\vec E}},\,\underline{{\vec B}}|
\{w_{ic}\,,\,w_{id}\}
 \, |W,\, \underline{{\vec E}},\,\underline{{\vec B}}\rangle.
\label{W11a1rr1} \ea
In this case the internal symmetry properties  are hard to  make
explicit. In order to determine whether or not the above contribution is zero we
contract (\ref{W11a1rr1}) with  the only three index tensor at our disposal:
$\epsilon^{a_1rr_1}$. The result is
\ba
\label{EPR}
\epsilon_{a_1rr_1}\, {R}_{11}{}^{a_1rr_1}&&=-\sum_{v \in \, {\rm Box}}\,  \frac{1}{2\,Q^2}\,
\left(\frac{2}{3!}\frac{8}{E(v)}\right)^2 \,
\sum_{v(\Delta)=v({\Delta'})=v}\frac{\mu^2}{6\,\ell_P^7}\,
det(s)\, \epsilon^{MNP}\, \nonumber \\
&& \times \,\langle W,\, \underline{{\vec E}},\, \underline{{\vec B}}|
\{w_{ic}\,,\,w_{id}\}\, |W,\, \underline{{\vec E}},\, \underline{{\vec B}} \rangle \,
\left(s'^d_P \, {s'}^c_M \,{s'}^q_N\, {s'}_{qM}\, + \frac{}{} \,
s'^d_P \,{s'}^c_N \,{s'}^q_M \, {s'}_{qM} \right). \nonumber \\
\ea
Inside the expectation value, upper spatial indices have been lowered by  the flat
metric. By symmetry requirements, the second term in parenthesis in (\ref{EPR}) yields zero.
Nevertheless, the first one gives the result
 \ba
{R}_{11}{}^{a_1rr_1}=\,\kappa_8\, \frac{\mu^2}{Q^2\,\ell_P^7}\, \ell_P^7\,
\frac{\ell_P^3}{{\cal L}^2}\, \left(\frac{\ell_P}{{\cal
L}}\right)^{2\Upsilon}\epsilon^{a_1rr_1}, \ea which produces a parity-violating term
in the magnetic sector of the effective Hamiltonian.

The next contribution arises from $T_2$ and  can be separated into three pieces \ba
\label{HM21}
{\rm H}_{21}^B&=&  \sum_{v \in V(\gamma)}\, \frac{1}{2\,Q^2}\, \left(\frac{2}{3!}\frac{8}{E(v)}\right)^2 \,\ell_P^3\, \sum_{v(\Delta)=v({\Delta'})=v} \,-\,\frac{1}{ \ell_P^7}\,\epsilon^{JKL}\, \epsilon^{MNP}\times \nonumber \\
&&\times \langle W,\, \underline{{\vec E}},\, \underline{{\vec B}}|\, U_{2LP}\,
(M_{1JK}\, M'_{3MN}+ M_{3JK}\, M'_{1MN}) \, |W,\, \underline{{\vec E}},\,
\underline{{\vec B}}\rangle, \ea \ba \label{HM22}
{\rm H}_{22}^B&=&  \sum_{v \in V(\gamma)}\, \frac{1}{2\,Q^2}\, \left(\frac{2}{3!}\frac{8}{E(v)}\right)^2 \,\ell_P^3\, \sum_{v(\Delta)=v({\Delta'})=v} \,-\,\frac{1}{\ell_P^7}\,\epsilon^{JKL}\, \epsilon^{MNP}\times \nonumber \\
&&\times \langle W,\, \underline{{\vec E}}\, \underline{{\vec B}}|\, U_{2LP}\,
M_{2JK} \, M'_{2MN}  \, |W,\, \underline{{\vec E}},\, \underline{{\vec B}}\rangle,
\ea \ba \label{HM23}
{\rm H}_{23}^B&=&  \sum_{v \in V(\gamma)}\, \frac{1}{2\,Q^2}\, \left(\frac{2}{3!}\frac{8}{E(v)}\right)^2 \,\ell_P^3\, \sum_{v(\Delta)=v({\Delta'})=v} \,-\,\frac{1}{\ell_P^7}\,\epsilon^{JKL}\, \epsilon^{MNP}\times \nonumber \\
&&\times \langle W,\, \underline{{\vec E}},\, \underline{{\vec B}}|\, U_{4LP}\,
M_{1JK}\, M'_{1MN} \, |W,\, \underline{{\vec E}},\, \underline{{\vec B}}\rangle. \ea

Let us start discussing  ${\rm H}_{21}^B$. After some algebra we obtain \ba
{\rm H}_{21}^B &=&  \sum_{v \in V(\gamma)}\, \frac{1}{2\,Q^2}\, \left(\frac{2}{3!}\frac{8}{E(v)}\right)^2 \,\ell_P^3\, \sum_{v(\Delta)=v({\Delta'})=v} \,-\,\frac{i}{\ell_P^7}\,\epsilon^{JKL}\, \epsilon^{MNP}\times \nonumber \\
&&\qquad \qquad \times \,\, s^a_L s'^d_P \, s'^x_N s'^q_M \, \langle W,\,
\underline{{\vec E}},\, \underline{{\vec B}}|\, w_{ia}w_{id} \,\left( s^r_K s^t_J
\frac{i}{4!} (x_J^2+x_Jx_K+x_K^2) \underline{F}_{rt}
 \right.\nonumber\\
&& \qquad \qquad\left.-s^r_K s^t_Js^u_K s^v_J\left[\frac{1}{4\cdot 3!} (x_J+x_K)\underline{F}_{rt}\underline{F}_{uv} + \frac{1}{4\cdot 3!} \underline{F}_{rt}(x_K+x_J)\underline{F}_{uv}\right] \right.\nonumber \\
&&\left.\qquad \quad \quad  - s^r_K s^t_J s^u_K s^v_J s^w_K s^z_J \frac{i}{4\cdot 3!}
\underline{F}_{rt} \underline{F}_{uv} \underline{F}_{wz} \,\right)
\underline{F}_{xq}\, |W,\, \underline{{\vec E}},\, \underline{{\vec B}} \rangle, \ea
which naturally splits into the following pieces
 \ba \label{H211} {\rm H}_{211}^B&=&
\sum_{v \in V(\gamma)}\, \frac{1}{2\,Q^2}\, \left(\frac{2}{3!}\frac{8}{E(v)}\right)^2 \,\ell_P^3\, \sum_{v(\Delta)=v({\Delta'})=v} \,-\,\frac{i}{ \ell_P^7}\,\epsilon^{JKL}\, \epsilon^{MNP}\times \nonumber \\
&&\times \, s^a_L s'^d_P \, s'^x_N s'^q_M \, \langle W,\, \underline{\vec{E}},\,
\underline{\vec{B}}|\, w_{ia}w_{id} \,\left(
s^r_K s^t_J \frac{i}{4!} (x_J^2+x_Jx_K+x_K^2)  \underline{F}_{rt}\right) \underline{F}_{xq}\, |W,\, \underline{\vec{E}},\, \underline{\vec{B}} \rangle,\nonumber\\
\ea \ba \label{H212}
{\rm H}^B_{212}&=&\sum_{v \in V(\gamma)}\, \frac{1}{2\,Q^2}\, \left(\frac{2}{3!}\frac{8}{E(v)}\right)^2 \,\ell_P^3\, \sum_{v(\Delta)=v({\Delta'})=v} \,\frac{i}{ \ell_P^7}\,\epsilon^{JKL}\, \epsilon^{MNP}\times \nonumber \\
&&\times \, s^a_L s'^d_P \, s'^x_N s'^q_M \, \langle W,\, \underline{\vec{E}},\,
\underline{\vec{B}}|\, w_{ia}w_{id} \,\left(
s^r_K s^t_Js^u_K s^v_J\left[\frac{1}{4\cdot 3!} (x_J+x_K)\underline{F}_{rt}\underline{F}_{uv} \right.\right.\nonumber\\
&&\mbox{} \left.\left. + \frac{1}{4\cdot 3!}
\underline{F}_{rt}(x_K+x_J)\underline{F}_{uv}\right] \,\right) \underline{F}_{xq}\,
|W,\, \underline{\vec{E}},\, \underline{\vec{B}} \rangle, \ea \ba \label{H213}
{\rm H}^B_{213}&=&\sum_{v \in V(\gamma)}\, \frac{1}{2\,Q^2}\, \left(\frac{2}{3!}\frac{8}{E(v)}\right)^2 \,\ell_P^3\, \sum_{v(\Delta)=v({\Delta'})=v} \,\frac{i}{ \ell_P^7}\,\epsilon^{JKL}\, \epsilon^{MNP}\times \nonumber \\
&&\times \, s^a_L s'^d_P \, s'^x_N s'^q_M \, \langle W,\, \underline{\vec{E}}\,
\underline{\vec{B}}|\, w_{ia}w_{id} \,\left(
 s^r_K s^t_J s^u_K s^v_J s^w_K s^z_J
\frac{i}{4\cdot 3!} \underline{F}_{rt} \underline{F}_{uv} \underline{F}_{wz}
\,\right) \underline{F}_{xq}\, |W,\, \underline{\vec{E}}\, \underline{\vec{B}}
\rangle.
\nonumber\\
\ea Next we calculate the corresponding $R$ tensors. The first one is
\begin{eqnarray}
{R}_{211}{}^{a_{1}a_{2}rr_{1}} &=&\sum_{v\in {\rm Box}%
}\, \frac{1}{2}\, \left(\frac{2}{3!}\frac{8}{E(v)}\right)^2 \,\,\,\sum_{v(\Delta )=v({\Delta ^{\prime }})=v}\frac{2\, \mu^2}{4!\,Q^2 \ell_P^7}\,\;det(s^{\prime })det(s)\nonumber \\
&&\epsilon _{NKJ}\left( s^{-1}\right)^{rN}\;\epsilon ^{JKL}\,\,s_{L}^{a}\,\left(
2s_{J}^{a_{1}}\,s_{J}^{a_{2}}+s_{J}^{a_{1}}\,s_{K}^{a_{2}}\right)\langle
W,\,\underline{\vec{E}},\,\underline{\vec{B}}|\,w_{ia}w_{i}^{\;r_{1}}|W,\,\underline{\vec{E}},\,\underline{
\vec{B}}\rangle, \nonumber \\
{R}_{211}{}^{a_{1}a_{2}rr_{1}} &=&\frac{\mu^2}{Q^2 \ell_P^7}\,
\ell_P^8\,\frac{\ell_P^3}{{\cal L}^2}\, \left( \frac{\ell_P}{{\cal
L}}\right)^{2\Upsilon} \left( \frac{}{} \kappa_6\,\delta^{a_1a_2}\, \delta^{r r_1} +
\kappa_7\left( \delta^{a_1r} \, \delta^{a_2r_1} + \delta^{a_2r} \, \delta^{a_1r_1}
\right)\right),
\end{eqnarray}
that  we have constructed using  the most general tensor with four indices, which is symmetrical
in $a_1$ and $a_2$. The terms proportional to $\kappa_7$  contribute with
${\underline B}_a\partial^a\partial^b {\underline B}_b$ to the effective Hamiltonian. Keeping this term will be useful to construct  the
electric sector later. Notice that  we have also introduced the inverse matrix $\left( s^{-1}\right)
_{m}^{N}\,$ such that
\begin{equation}
\left( s^{-1}\right) _{m}^{P}\,  s_{Q}^{m}=\delta^P_Q,\qquad  s_{M}^{a}\,\left(
s^{-1}\right) _{b}^{M}=\delta^a_b,
\end{equation}
which elements scale as $\ell_P^{-1}$.

Our next contribution, arising from (\ref{H212}), is zero because the object (no sum
over $J$ and $K$) \be s^r_K s^t_Js^u_K s^v_J\, \underline{F}_{uv} \,
(x_J+x_K)\underline{F}_{rt},\ee which is symmetrical in $J\,K$, appears contracted
with $\epsilon^{JKL}$.

Now we consider
\begin{eqnarray}
{\rm H}_{213}^{B} &=&\sum_{{\rm Box}({\vec{x}})}\underline{B}^{b}({\vec{x}})\,\underline{B}^{c}({\vec{x}})\,\underline{B}^{f}({\vec{x}})\,\underline{B}^{e}({\vec{x}})%
\,\sum_{v\in Box}\,\frac{1}{2\,Q^2}\,\left( \frac{2}{3!}\frac{8%
}{E(v)}\right) ^{2}\,\ell _{P}^{3}\,\sum_{v(\Delta )=v({\Delta ^{\prime }}%
)=v}\,-\frac{\mu ^{4}}{4\times 3!\,\ell _{P}^{7}}\,  \nonumber \\
&&2\;det(s^{\prime })\left( det(s)\,\right) ^{3}\epsilon ^{JKL}\,s_{L}^{a}\,\left(
\epsilon _{UKJ}\left( s^{-1}\right) _{b}^{U}\right) \left( \epsilon _{VKJ}\left(
s^{-1}\right) _{c}^{V}\right)
\left( \epsilon _{XKJ}\left( s^{-1}\right) _{f}^{X}\right)\, \times\nonumber\\
&&\times \, \langle W,\,\underline{{\vec E}},\,\underline{{\vec
B}}|\,w_{ia}w_{ie}^{\;}\,|W,\,\underline{{\vec E}},\,\underline{{\vec B}}\rangle,
\end{eqnarray}
where we have introduced the relation \be \epsilon
_{mbc}\,\,s_{J}^{b}\,s_{L}^{c}=det(s)\,\epsilon _{KJL}\left( s^{-1}\right) _{m}^{K},
\ee obtained from (\ref{DET}). The further properties
\begin{eqnarray}
\epsilon ^{LKJ}\,\epsilon _{UKJ}\epsilon _{VKJ}\epsilon _{XKJ} \Longrightarrow
L=U=V=X, \qquad \epsilon ^{1KJ}\,\epsilon _{1KJ}\epsilon _{1KJ}\epsilon _{1KJ}
&=&2\epsilon
^{123}\,\epsilon _{123}\epsilon _{123}\epsilon _{123}=-2,\nonumber \\
\end{eqnarray}
allow us to rewrite the corresponding tensor in the simpler form
\ba
R_{213}{}_{a_1a_2a_3a_4}=&&\sum_{v \in {\rm Box}}\, \frac{1}{2\,Q^2}\, \left(\frac{2}{3!}\frac{8}{E(v)}\right)^2  \sum_{v(\Delta)=v({\Delta'})=v} \,-\frac{\mu^4}{3!\, \ell_P^7}\, \;det(s^{\prime })\left( det(s)\,\right) ^{3}\times \nonumber \\
&& \times \,s_{L}^{a}\,\left( s^{-1}\right) _{a_1}^{L}\left( s^{-1}\right)
_{a_2}^{L}\left( s^{-1}\right) _{a_3}^{L}\,\,\,\,\langle W,\,\underline{{\vec
E}},\,\underline{{\vec B}}|\,w_{ia}w_{ia_4}^{\;}\,|W,\,\underline{{\vec
E}},\,\underline{{\vec B}}\rangle. \ea
This shows  explicitly the symmetry in the
indices $a_1, \, a_2,\,  a_3$ and  leads to
\ba {R}_{213}{}_{a_1a_2a_3a_4}=\,\frac{\mu^4}{Q^2 \ell_P^7}\,
\ell_P^{10}\frac{\ell_P^3}{{\cal L}^2}\,\left(\frac{\ell_P}{{\cal
L}}\right)^{2\Upsilon}\,\kappa_4\, \left(\frac{}{}\delta_{a_1a_2}\, \delta_{a_3a_4} +
\delta_{a_1a_3}\, \delta_{a_2a_4}+ \delta_{a_1a_4}\, \delta_{a_2a_3}
\right),\label{W213} \ea where we have written the most general  four index tensor at
our disposal, which is  completely symmetric in three indices. This implies the
complete symmetry in the four indices. The correction (\ref{W213}) leads to cubic
non-linear terms in the equation of motion.

Now we continue with the correction arising from (\ref{HM22}), which reduces to  \ba {\rm H}_{22}^B&=&
\sum_{v \in {\rm Box}}\, \frac{1}{2\,Q^2}\, \left(\frac{2}{3!}\frac{8}{E(v)}\right)^2 \,\ell_P^3\, \sum_{v(\Delta)=v({\Delta'})=v} \,\,\frac{4}{(3!)^2\,\ell_P^7}\,\epsilon^{JKL}\, \epsilon^{MNP}\times \nonumber \\
&&\times s^a_L s'^d_P\, \langle W,\, \underline{\vec{E}},\, \underline{\vec{B}}|\,
w_{ia}w_{id}\,\left(s^u_K s^y_J\,x_J \underline{F}_{uy}\right)\, \left(s'^r_N
s'^s_M\, x'_M \underline{F}_{rs} \right) \, |W,\, \underline{\vec{E}},\,
\underline{\vec{B}} \rangle,
\ea
because the terms quadratic in $\underline{F}$ are symmetric in $J,K$
and $M,N$ respectively.

>From the above we read
\ba
{R}_{22}{}^{a_1a_2rr_1} &=& \sum_{v \in {\rm Box}}\, \frac{1}{2\,Q^2}\,
\left(\frac{2}{3!}\frac{8}{E(v)}\right)^2 \sum_{v(\Delta)=v({\Delta'})=v} \,\,\frac{4\,
\mu^2}{(3!)^2\, \ell_P^7}\, \,det(s)det(s^{\prime
})\times \nonumber \\
&&\times \left( -\epsilon _{YMN}\epsilon ^{PMN}s^{\prime }{}_{P}^{d}\;s^{\prime
}{}_{M}^{a_{2}}\right) \;\left( \epsilon _{XJK}\epsilon
^{LJK}\,\,s_{L}^{a}\,\,\,s_{J}^{a_{1}}\right) \;\,\left( s^{-1}\right) ^{Xr_{1}}\,\left( s^{\prime -1}\right) ^{Yr}\nonumber \\
&&\times \, \langle W,\,\underline{{\vec E}},\,\underline{{\vec
B}}|\,w_{ia}w_{id}\,|W,\,\underline{{\vec E}},\,\underline{{\vec B}}\rangle. \ea

Here the internal symmetry properties are rather obscure. Nevertheless, the symmetry
properties induced by  the classical magnetic field factors imply that the above
tensor must be proportional to the most general tensor with four indices which is
symmetric in $a_1\, a_2$. Then we have
\ba R_{22}{}^{a_1a_2rr_1} &=& \frac{
\mu^2}{Q^2 \ell_P^7}\, \ell_P^8\,\frac{\ell_P^3}{{\cal L}^2}\,
\left(\frac{\ell_P}{{\cal L}}\right)^{2\Upsilon}\left( \frac{}{}
\kappa_9\,\delta^{a_1a_2}\, \delta^{r r_1} + \kappa_{10}\left( \delta^{a_1r} \,
\delta^{a_2r_1} + \delta^{a_2r} \, \delta^{a_1r_1}  \right)\right). \ea This
contribution is of the same kind as the one given by ${R}_{211}{}^{a_1a_2rr_1}$.

Finally we are left with
\ba
{\rm H}_{23}^B
&=& \sum_{{\rm Box}({\vec x})}\underline{B}_{r_1}({\vec x})\,\underline{B}_r({\vec x})\, \sum_{v \in {\rm Box}({\vec x})}\, \frac{1}{2\,Q^2}\, \left(\frac{2}{3!}\frac{8}{E(v)}\right)^2 \,\ell_P^3\, \sum_{v(\Delta)=v({\Delta'})=v} \,\frac{\mu^2}{4\, \ell_P^7}\, \times \nonumber \\
&&\times \,\epsilon^{JKL}\, \epsilon^{MNP}\, s^u_K s^v_J \, s'^x_N s'^y_M \,
\left(s^d_L s'^a_P s'^b_P s'^c_P\right) \, \epsilon_{uv}{}^{r_1}\, \epsilon_{xy}{}^z
\langle W,\, \underline{{\vec E}},\, \underline{{\vec B}}|\, \{ w_{id}, \, w_{iabc}\} \,  |W,\, \underline{{\vec E}},\, \underline{{\vec B}} \rangle, \nonumber \\
\ea
which leads to
\begin{eqnarray}
R_{23}{}^{rr_1}&=&\sum_{v\in {\rm Box}({\vec{x}})}\,\frac{1}{%
2\,Q^2}\,\left( \frac{2}{3!} \frac{8}{E(v)}\right) ^{2}\,\sum_{v(\Delta )=v({\Delta ^{\prime }})=v}\,-\frac{\mu ^{2}}{\ell _{P}^{7}}\, \det (s^{\prime })\, det(s)\nonumber \\
&&\times \;\left( s^{\prime -1}\right)
^{rP}\,\left( {s^{\prime }}_{P}^{a}\;{s^{\prime }}_{P}^{b}\;{s^{\prime }}%
_{P}^{c}\,\;\right) \;\langle W,\,\underline{{\vec E}},\, \underline{{\vec
B}}|\,\{w_{i}^{\;r_{1}},\,w_{iabc}\}\,|W,\,\underline{{\vec E}},\, \underline{{\vec
B}}\rangle.
\end{eqnarray}
Taking the symmetric part, we have
\begin{eqnarray}
R_{23}{}^{rr_1}&=&\kappa_{11} \, \frac{\mu ^{2}}{%
Q^{2}\ell _{P}^{7}}\,\ell_P^8\,\frac{\ell_P^3}{{\cal L}^4}\,\left( \frac{\ell_P}
{{\cal L}}\right)^{4\Upsilon}\delta^{r\,r_1}.
\end{eqnarray}

Adding all previous contributions, we obtain the magnetic sector of the effective
Hamiltonian, up to order $\ell_P^2$,
\ba \label{HMFIN} {\rm H}^B&=& \frac{1}{Q^2}\int
d^3({\vec x}) \left[\left(1+ \theta_7 \,\left(\frac{\ell_P}{{\cal
L}}\right)^{2+2\Upsilon} \right)\frac{1}{2}\underline{{\vec B}}^2+ \ell_P^2
\,\left(\frac{}{}\theta_2\,{\underline B}^a \partial_a \partial_b {\underline B}^b  +
\theta_3 \,  \underline{B}^a \, \nabla^2 \underline{B}_a \right)+
\ \right. \nonumber \\
&&\left. \qquad \qquad \qquad+ \theta_8 \ell_P \underline{\vec B}\cdot(\nabla
\times\underline{\vec B})+ \theta_4\, {\cal L}^2\, \ell_P^2 \,\left(\frac{{\cal
L}}{\ell_P} \right)^{2\Upsilon}\left(\underline{{\vec B}}^2\frac{}{}\right)^2+\dots
\right]. \ea
The numbers $\theta_i$ are linear combinations of the
corresponding $\kappa_j$ appearing in the tensors $R$. The correspondences are \ba
\kappa_7, \kappa_{10} \rightarrow \theta_2, \quad \kappa_6, \kappa_{9} \rightarrow
\theta_3,\quad \kappa_4 \rightarrow \theta_4,\quad \kappa_{11} \rightarrow
\theta_7,\quad, \kappa_8 \rightarrow \theta_8. \ea
As it is pointed out  after (\ref{EHC}), the electric sector of the effective Hamiltonian can be obtained by changing ${\vec B}$ into $\vec {E}$ in the quadratic contribution of the corresponding magnetic sector. In this way the complete
electromagnetic effective Hamiltonian becomes
\ba \label{HEMFIN} {\rm H}^{EM}&=& \frac{1}{Q^2}\int d^3{\vec x} \left[\left(1+
\theta_7 \,\left(\frac{\ell_P}{{\cal L}}\right)^{2+2\Upsilon}
\right)\frac{1}{2}\left(\frac{}{} \underline{{\vec B}}^2 + \underline{{\vec
E}}^2\right) + \theta_3 \, \ell_P^2 \, \left( \frac{}{}\underline{B}^a \,\nabla^2
\underline{B}_a + \underline{E}^a \,\nabla^2 \underline{E}_a\right)+
\ \right. \nonumber \\
&&\left. \qquad \qquad \qquad+ \theta_2\,\ell_P^2\,{\underline E}^a
\partial_a \partial_b {\underline E}^b+
\theta_8 \ell_P \left( \frac{}{} \underline{\vec B}\cdot(\nabla \times\underline{\vec
B})+
\underline{\vec E}\cdot(\nabla \times\underline{\vec
E}) \right)\right.\nonumber \\
&&\left. \qquad \qquad \qquad+ \theta_4\, {\cal L}^2 \, \ell_P^2 \, \left(\frac{{\cal
L}}{\ell_P} \right)^{2 \Upsilon}\, \left(\frac{}{}\underline{{\vec B}}^2\right)^2
+\dots \right], \ea
up to order $\ell_P^2$.

\section{Modified Maxwell equations and dispersion relations in vacuum}

\label{MMEQ}

Since no confusion arises in the sequel, we eliminate the underline in all
electromagnetic quantities. From the  effective Hamiltonian
(\ref{HEMFIN}) we obtain the equations of motion
\begin{eqnarray}
A (\nabla\times\vec B)-\frac{1}{c}\frac{\partial\vec E}{\partial t}%
+ 2\ell_P^2\theta_3\nabla^2(\nabla\times {\vec B}) -2\theta_8\ell_P\nabla^2\vec
B +4\theta_4{\cal L}^2\left(\frac{\cal L}{\ell_P}\right)^{2\Upsilon}
\ell_P^2\nabla\times(\vec B^2\vec B)=0,\label{ME1} \\
A (\nabla\times\vec E)+\frac{1}{c}\frac{\partial\vec B}{\partial t}+
2\ell_P^2\theta_3\nabla^2(\nabla\times {\vec E})-2\theta_8\ell_P\nabla^2\vec E=0,
\label{ME2}
\end{eqnarray}
where
\begin{eqnarray}
A=1+\theta_7\left(\frac{\ell_P}{{\cal L} }\right)^{2+2\Upsilon}.
\end{eqnarray}
The above equations are supplemented by the condition $\nabla\cdot{\vec B}=0$%
, together with the constraint $\nabla\cdot{\vec E}=0$, appropriate for vacuum.

Next we calculate the dispersion relations arising from the modified Maxwell equations (\ref{ME1}) and (\ref{ME2}).
Neglecting the non-linear part and introducing the plane wave ansatz
\begin{eqnarray}
\vec E=\vec E_0\,e^{i(\vec k\vec x-\omega t)},\qquad \vec B=\vec B_0\, e^{i(\vec
k\vec x-\omega t)},\qquad k=|\vec k |,
\end{eqnarray}
we get
\begin{eqnarray}
\vec E_0\cdot \vec k=0, \qquad \vec B_0 \cdot\vec k=0, \\
 (\vec k\times\vec
E_0)\left[A-2\,\theta_3\,(\ell_P\,{k})^2\right]-2\,i\,\theta_8%
\ell_P\, {\ k}^2\vec E_0-\frac{\omega}{c}\vec B_0=0,
\\
(\vec k\times\vec B_0)\left[A-2\,\theta_3\,(\ell_P\,{\ k})^2\right]-2\,i\,\theta_8%
\ell_P\, {\ k}^2\vec B_0+\frac{\omega}{c}\vec E_0= 0,
\end{eqnarray}
which imply the following dispersion relations
\begin{eqnarray}
\omega=c k\left(1+\theta_7\left(\frac{\ell_P}{{\cal L} }\right)^{2+2\Upsilon}-2\,\theta_3\,(k\ell_P)^2\pm 2\theta_8\,(k\ell_P ) \right).
\end{eqnarray}
The $\pm$ signs correspond to the different polarizations of the photon. From the above we
obtain the speed of the photon

\begin{eqnarray}
\frac{v}{c}=\frac{1}{c}\, \frac{d\omega}{d k}\left|\frac{}{}\right._{{\cal L}%
=1/k}=1 \pm 4\, \theta_8 \,(k \ell_P)  -6\theta_3 (k \ell_P)^2+\theta_7\,\left(k{\ell_P}\right)^{2+2\Upsilon}+
... \label{PHOTV}
\end{eqnarray}

The last expression gives $v$ expanded to leading order in $\ell_P$ where we have estimated ${\cal L}$ as $%
1/k$, which is its maximal value. Clearly (\ref%
{PHOTV}) is valid only for momenta satisfying $(\ell_P\,k) <\!< 1$.

To first order in $(k \ell_P)$  we have  only  the helicity dependent correction found already by Gambini and Pullin \cite{GP}. As far as the $\Upsilon$ dependent terms we have either a quadratic ($\Upsilon=0$) or a quartic ($\Upsilon=1$) correction. The only possibility to have
a first order helicity independent correction amounts to set $\Upsilon=-1/2$ which
corresponds to that of Ellis et. al. \cite{ELLISETAL}. However, we do not have an
interpretation for such a value of $\Upsilon$.

\section{Effect of non-linear terms}

\label{NLT}

In this section we explore some  implications of the non-linear term in the
Maxwell equations  induced by the quantum gravity corrections to order
$\ell_P^2$.

Following reference \cite{Landau} we study the propagation of waves in the presence
of a constant magnetic field $\vec B_0$. To do this, let us write
\begin{equation}
\vec B=\vec B_0 +\vec b
\end{equation}
and consider only the contribution of the non-linear term in the effective Hamiltonian. Namely, let us set  $%
\theta_3=\theta_7=\theta_8=0$. After linearizing
in $\vec b$, the field equations reduce to
\begin{eqnarray}
\nabla\times\vec b-\frac{1}{c}\frac{\partial\vec E}{\partial t}+
4\,\bar\theta_4\,{\cal L}^2 \ell_P^2\, \nabla\times(\vec B_0^2\vec b+2(\vec
B_0\cdot\vec b)\vec B_0)&=&0,  \nonumber \\
{\nabla}\cdot {\vec E}=0, \qquad {\nabla}\cdot {\vec b}=0,\qquad
\nabla\times\vec E+\frac{1}{c}\frac{\partial\vec b}{\partial t}&=&0,
\label{NLME}
\end{eqnarray}
with $\bar\theta_4=\left(\frac{\cal L}{\ell_P}\right)^{2\Upsilon}\theta_4.$
Now we look for plane wave solutions of (\ref{NLME}) with
\begin{eqnarray}
\vec E=\vec E_0\,e^{i(\vec k\vec x-\omega t)},\qquad \vec b=\vec b_0\,e^{i(\vec k\vec
x-\omega t)},
\end{eqnarray}
obtaining the conditions
\begin{eqnarray}
\vec E_0\cdot \vec k=0, \qquad \vec b_0 \cdot\vec k=0, \qquad (\vec
k\times\vec E_0)-\frac{\omega}{c}\vec b_0= 0, \label{NONLE1} \\
(1+4\,\bar\theta_4\,{\cal L}^2\ell_P^2\,\vec B_0 ^2 )(\vec k\times\vec b_0)+%
\frac{\omega}{c}\vec E_0+ 8\,\bar\theta_4\,{\cal L}^2\ell_P^2\,(\vec b_0\cdot\vec
B_0)(\vec k\times\vec B_0) = 0.  \label{NONLE}
\end{eqnarray}
Substituting in (\ref{NONLE}) the expression for $\vec b_0$ obtained from the third equation in (\ref{NONLE1})
 we are left with
\begin{equation}
\left(\frac{\omega^2}{c^2}-k^{2} \left( 1+4\,\bar\theta _{4}\,{\cal L}^{2}\ell
_{P}^{2}\,\vec{B}_{0}^{2}\right)\right)\vec{E}_{0}=- 8\,\bar\theta _{4}\,{\cal L}%
^{2}\ell _{P}^{2}\,\left(\frac{}{}\left( \vec{k} \times \vec{E}_{0}\right) \cdot
\vec{B}_{0})\right)(\vec{k}\times \vec{B}_{0}).  \label{FDR}
\end{equation}
Since $\vec B_0$ and $\vec k$ determine a plane, it is natural to study separately
the propagation of waves with polarization parallel and perpendicular to this plane.
We will express the answer in terms of the refraction index
\begin{equation}
n=\frac{k\,c}{\omega}.
\end{equation}
>From (\ref{FDR}) we obtain the following refraction indices ($\hbar=1=c$)
\begin{eqnarray}
n_{\parallel}&=&1-2\bar\theta_4\,\frac{1}{\omega^2}\, \ell_P^2 \,\vec B_0^2,
\nonumber \\
n_{\perp}&=&1-2\bar\theta_4\,\frac{1}{\omega^2}\,\ell_P^2 \,\vec B_0^2\,(1+2\, {\rm
sin}^2\phi),
\end{eqnarray}
for parallel and perpendicular photon polarization, respectively. Here $\phi$ is the
angle between $\vec k$ and $\vec B_0$.

These results can be compared with similar effects in quantum electrodynamics
\cite{Landau}
\begin{eqnarray}
n_{\parallel}=1+\theta_{\parallel}\vec B_0^2\,{\rm sin}^2\phi,  \nonumber \\
n_{\perp}=1+\theta_{\perp}\vec B_0^2\,{\rm sin}^2\phi.
\end{eqnarray}
where
\begin{equation}
\theta_{\parallel}=\frac{2e^4}{45m^4}, \qquad \theta_{\perp}=\frac{7e^4}{%
90m^4}.
\end{equation}

As expected, quantum gravity induced effects are much smaller than purely quantum
electrodynamics effects, but the former present different signatures. In particular,
the indices arising from quantum gravity are frequency dependent and also
$n_{\parallel}$ is independent of $\phi$.

\section{Photon time delay}

\label{SPHOTD}

Notice that our considerations assumed a coarse-grained flat spacetime rather than a
FLRW model. The latter would seem more appropriate for GRB traveling cosmological
distances. In particular, it would be interesting to study the redshift effects in the photon time delays induced by the energy dependent corrections to the velocity.
We are able to estimate these effects as
follows.

Using the flat FLRW metric let us calculate the present time delay of two photons emitted simultaneously with
different momenta and hence different velocities. We set $r=r_{1}=L,\,t=t_{1}$ as the emission coordinates and $%
r=0,t=t_{0}$ as the detection coordinates in the comoving cosmological system $%
t,r,\theta(t)=0 $. The definition of the velocity $V(t)=
R(t)\,\frac{d\,r}{d\,t}$, with $\frac{d\,r}{d\,t}<0$,
leads to
\begin{equation}
r_{1}=\int_0^{r_1} d\,r=\int_{t_{1}}^{t_{0}}\frac{V(t)}{R(t)}\,dt.
\end{equation}
Here $R(t)$ is the scale factor in the FLRW metric. Notice that the above equation leads to the standard red shift result for  photons moving  with
$V=c$. On the other hand, quantum gravity corrections predict
\begin{equation}  \label{QGCFIN}
\frac{V}{c}= 1 + \theta\, (\ell_P k)
\end{equation}
to leading order.
We are interested in discussing the situation where two photons are emitted at $r_1$
with different velocities $V_{1}$ and $V_{1}+\delta \,V_{1}$, and arrive at $r=0$ at
times $t_{0}$ and $t_{0}+\delta t_{0}$ respectively. Then we have
\begin{equation}
r_{1}=\,\int_{t_{1}}^{t_{0}}\,\frac{d\,t}{R(t)}\,V(t),\qquad
r_{1}=\,\int_{t_{1}}^{t_{0}+\delta \,t_{0}}\,\frac{d\,t}{R(t)}\left( V(t)+\delta
\,V(t)\frac{{}}{{}}\right).
\end{equation}
Subtracting the two expressions for the fixed coordinate $r_{1}$, we obtain
\begin{equation}
\delta \,t_{0}=-\frac{R_{0}}{V(0)}\,\int_{t_{1}}^{t_{0}}\,\frac{d\,t}{R(t)}%
\,\delta \,V(t).
\end{equation}
In the flat FLRW universe we have
\begin{equation}
t=\frac{t_{0}}{(1+z)^{3/2}},\quad t_{0}=\frac{2}{3}\,\frac{1}{H_{0}},\quad
dt=-\frac{1}{H_{0}}\,\frac{1}{(1+z)^{5/2}}\,d\,z,\quad \frac{R(t)}{R_{0}}=%
\frac{1}{1+z},
\end{equation}
where $z\,$ is the red shift. This leads to
\begin{equation}  \label{PHOTD}
\delta \,t_{0}=\frac{1}{c\,H_{0}}\int_{0}^{z_{1}}\frac{d\,z}{(1+z)^{3/2}}%
\,\delta \,V(z),
\end{equation}%
where $z_{1}$ is the red shift of the source and $\delta\, V(z)$ is the difference
between the velocities of the two photons, arising from the quantum gravity
corrections (\ref{QGCFIN}). Using the zeroth-order relation for the red shift of the
photon momentum,
\begin{equation}
k(z)=\frac{R_0}{R(z)}k_0,
\end{equation}
we obtain
\begin{equation}
\delta V(z)=\theta\,\frac{R_0}{R(z)}\;(\delta k)_0\,\ell _{P}= \theta\,(1+z)\;
(\delta k)_0\,\ell _{P},
\end{equation}
where $(\delta k)_0$ is the present day  observed momentum difference between the two
photons.

Substitution of the above equation in (\ref{PHOTD}) yields
\begin{equation}
\delta \,t_{0} = \left( \frac{2\,\theta\;}{H_{0}}\right)\,(\delta k)_0\, \ell
_{P}\,\left( (1+z_1)^{1/2}-1\right).
\end{equation}
The above result differs from the corresponding one obtained in the second and third
papers of Ref. \cite{ELLISETAL}.

\section{Summary and Discussion}

\label{disc}

In this paper we have considered the propagation of photons in a semiclassical background provided
by loop quantum gravity. An effective electromagnetic Hamiltonian, given by  (\ref{HEMFIN}),
was identified with the expectation value of the electromagnetic piece of the
Hamiltonian constraint for the Einstein-Maxwell theory with respect to {\it a would be
semiclassical state}. The state used was assumed to approximate a classical flat
metric, a classical flat gravitational connection and a generic classical
electromagnetic field, at scales larger than the coarse-grained characteristic length
${\cal L}$, where ${\cal L}>\!>\ell_P$ (the Planck length). To leading
non trivial order in $\ell_P$, photons of wavelength $\lambda$, where ${\cal %
L} < \lambda$, acquire Planck scale modified dispersion relations as compared to
those in classical flat spacetime. This in turn yields the effective speed of
light  (\ref{PHOTV})
which involves two types of corrections. One of them is just that of Gambini and
Pullin \cite{GP} including the helicity  of the photon, whereas the other
depends on the scale $\cal L$. Moreover, the latter type contains
a parameter $\Upsilon$ that encodes the scaling of the
gravitational connection under the semiclassical expectation value.

When estimating the coarse-grained characteristic length by ${\cal L}=1/k$,
which is its maximal value, the following values of $\Upsilon$ are prominent:
(i) $\Upsilon=0$ can be understood as that the connection can not be probed below the
coarse graining scale $\cal L$. The corresponding correction scales as $(k\ell_P)^2$.
(ii) $\Upsilon=1$ may be interpreted as the analog of
the coherent states analysis, where such states saturate the Heisenberg uncertainty relation
inside a box of volume ${\cal L}^3$: $\Delta q \sim \frac{\ell_P}{\cal L}$, $\Delta
A\sim \frac{\ell_P}{{\cal L}^2}$ and $\Delta q \Delta A \sim \frac{\kappa \hbar}{{\cal
L}^3}$ \cite{HEITLER}. Then the correction behaves like $(k\ell_P)^4$. (iii) Interestingly, a value
$\Upsilon=-\frac{1}{2}$ leads to a helicity independent first order
correction (i.e. $(k\ell_P)$) similar to that of Ellis et al. \cite{ELLISETAL}.
We do not have an interpretation of this case though.

A prime candidate for testing the  effects  which are linear in the energy
would be the Gamma Ray Bursts that travel
cosmological distances and which might be detected with a time resolution beyond $%
10^{-5}$ seconds. This seems possible in future spatial experiments \cite%
{meszaros}.

Moreover, new non-linear terms in the Maxwell equations appear. These terms are not
present either in \cite{GP} or \cite{ELLISETAL}. We have explored the significance of
this contribution to the propagation of photons in a constant strong magnetic field.
The corrections obtained in the corresponding refraction indices are much
smaller than similar effects in Quantum Electrodynamics. Nevertheless, quantum
gravity corrections have distinct signatures: a main difference is that the speed of
photons with polarization parallel to the plane formed by the background magnetic
field and the direction of the wave is isotropic.

Our results should be taken as first steps in the exploration of possible observable
consequences of quantum gravity. We have given evidence that dispersion relations of
the form (\ref{eq:drel}) can  have origin in the microstructure of spacetime. It is
expected that the recently proposed coherent states for quantum gravity and gauge
theories in  \cite {aei} and/or statistical geometry \cite{statg}, will come to
terms systematically with the unknown numerical coefficients we have left
undetermined in our calculation (See also \cite{madzap}.) Interestingly, a quantum
field theory in effective spacetimes might possibly emerge along these lines. Yet
another avenue in the context of canonical quantization of gravity and gauge theories
has recently emerged to understand the semiclassical regime. It is
aimed at establishing a relation between Fock space and ${\cal
H}_{\mathrm aux}$ \cite{FockU1,Fockpolymer}.

Further work remains to be done in the framework here developed. For instance, in the
case of inflationary cosmology as well as in the study of the Hawking effect, use is
made of scalar fields with non standard near Planckian frequency dispersion relations
to model the effect of short distance physics on the quantum fields \cite{jacobson}. Indeed a
systematic study of the modifications induced by quantum gravity along the lines we have
developed could be performed to investigate whether the dispersion relations used in
\cite{jacobson} can be accounted for.

Finally, we stress that the dispersion relations we have found, as well as those in \cite{GP,ELLISETAL,URRU1,ELLISFERM,jacobson} are
Lorentz symmetry violating. This is not necessarily an issue as it has been
extensively discussed previously \cite{ACPol,ACEssay}. In fact they may alleviate
some long standing astrophysical and cosmological problems \cite{AMC4,ALF}. Remarkably,
there has been  considerable progress in setting bounds to Lorentz invariance violation \cite{kostelecky, tritium,liberati}
and to the values of some coefficients in the effective Maxwell equations \cite{gleiser}.

\

{\bf Acknowledgments}. The authors are grateful to A. Ashtekar, M. Bojowald, R.
Gambini and T. Thiemann for illuminating comments at different stages of this work.
Partial support is acknowledged from CONICyT-CONACyT E120-2639, DGAPA IN11700 and
CONACYT 32431-E. We also acknowledge the project Fondecyt 7010967.
 The work of JA is partially supported by Fondecyt 1010967. HAMT thanks for the warm hospitality and
stimulating environment experienced at the Center for Gravitational Physics and
Geometry-PSU, in particular to A. Ashtekar and J. Pullin. He also acknowledges
partial support from NSF Grants PHY00-90091 and  INT97-22514, Eberly Research Funds at Penn State,
CONACyT 55751 and M\'exico-USA Foundation for
Science. LFU acknowledges the hospitality at the Center for Gravitational Physics and
Geometry-PSU.



\begin{thebibliography}{99}
\bibitem{AC} G. Amelino-Camelia, J. Ellis, N.E. Mavromatos, D.V. Nanopoulos
and S. Sarkar, Tests of quantum gravity from observations of $\gamma$-ray bursts, {
Nature}\, {\bf 393} \,(1998)\, 763-765.

\bibitem{ELLIS} J. Ellis, N.E. Mavromatos and D.V. Nanopoulos, Search for quantum gravity,
{ Gen. Rel. Grav.} {\bf 31} (1999) 1257 [gr-qc/9905048].

\bibitem{AHLU} D.V. Ahluwalia, Quantum gravity: testing times for theories,
{ Nature} {\bf 398} (1999) 199; G.Z. Adunas, E. Rodriguez-Milla and D.V. Ahluwalia,
Probing quantum aspects of gravity, { Phys. Letts.} {\bf B485}(2000)215-223,
[gr-qc/0006021]; G.Z. Adunas, E. Rodriguez-Milla and D.V. Ahluwalia, Probing quantum
violations of the equivalence principle, { Gen. Rel. Grav.} {\bf
33}(2001)183-194,[gr-qc/0006022]; D.V. Ahluwalia, C.A. Ortiz and G.Z. Adunas, {
Robust flavor equalization of cosmic neutrino flux by quasi bimaximal mixing},
[hep-ph/0006092].

\bibitem{ACPol}
G. Amelino-Camelia,  { Are we at the dawn of quantum gravity phenomenology?},
Lectures given at 35th Winter School of Theoretical Physics: From Cosmology to
Quantum Gravity, Polanica, Poland, 2-12 Feb 1999. Published in Lect. Notes Phys. 541
(2000) 1-49, Springer Verlag  [gr-qc/9910089].

\bibitem{ACEssay}
G. Amelino-Camelia, Planck length phenomenology, { Int. Jour. Mod. Phys.} {\bf D10}
(2001) 1, [gr-qc/0008010].

\bibitem{METZ} J. van Paradis et. al., Transient optical emission from the error
box of the $\gamma$-ray burst of 28 february 1997, { Nature} {\bf 386}(1997)686; P.J.
Groot et al.,IAU Circ. No 6676, 1997; M. L. Metzger et al.,  Spectral constraints  on
the red shift  of the optical counterpart of the $\gamma$-ray burst of 8 May 1997, {
Nature} {\bf 387 }(1997)878.

\bibitem{BHAT} P.N. Bhat, G.J. Fishman, C.A. Meegan, R.B. Wilson, M.N. Brock
and W.S. Paclesas, Evidence for sub-millisecond structure in a $\gamma$-ray burst, {
Nature} {\bf 359}(1992)217.

\bibitem{meszaros} P. M\'esz\'aros, Gamma-Ray bursts
and bursters, { Nucl. Phys. B (Proc. Suppl.)} {\bf 80} (2000) 63-77.


\bibitem{GP} R. Gambini and J. Pullin, Non standard optics from quantum spacetime,
{\em Phys. Rev.} {\bf D59} (1999) 124021, [gr-qc/9809038].

\bibitem{ELLISETAL} J. Ellis, N.E. Mavromatos and D.V. Nanopoulos, Quantum gravitational
difussion and stochastic fluctuations in the velocity of light, {\em Gen. Rel. Grav.}
{\bf 32} (2000) 127-144, [gr-qc/9904068]; J. Ellis, N.E. Mavromatos and D.V.
Nanopoulos, Tegernsee 1999, Beyond the desert, p.299-334,[gr-qc/9909085] and
references therein; J. Ellis, K. Farakos, N.E. Mavromatos, V.A. Mitsou and D.V.
Nanopoulos, Astrophysical Probes of the Constancy of the Velocity of Light, {\em
Astrophysical Jour.} {\bf 535} (2000) 139-151, [astro-ph/9907340].

\bibitem{REVIEW} N.E. Mavromatos, { The quest for quantum gravity:
testing times for theories?}, [astro-ph/0004225]; J. Ellis,{ Perspectives in
High-Energy Physics}, JHEP Proceedings, SILAFAE III, Cartagena de Indias, Colombia,
april 2-8,2000, [hep-ph/0007161]; J. Ellis, {\it Testing fundamental physics with
high-energy cosmic rays}, astro-ph/0010474.

\bibitem{oneloop-effqg}
D.A.R. Dalvit, F.D. Mazzitelli, C. Molina-Paris, One loop graviton corrections to
Maxwell's equations, { Phys. Rev.} {\bf D63}(2001) 084023, [hep-th/0010229].

\bibitem{opensystems}
F. Benatti and  R. Floreanini, Massless neutrino oscillations,  [hep-ph/0105303];
Effective dissipative dynamics for polarized photons, { Phys. Rev.} {\bf D62} (2000)
125009,    [hep-ph/0009283].

\bibitem{Ford99}
H. Yu and  L.H. Ford, Lightcone fluctuations in flat spacetimes with non trivial topology,
Phys. Rev. {\bf D60} (1999) 084023, [gr-qc/9904082].

\bibitem{WAX} E. Waxman and J. Bahcall, High energy neutrinos from cosmological gamma
ray burst fireballs, { Phys. Rev. Letts.} {\bf 78}(1997) 2292-2295; E. Waxman, High
energy neutrinos from gamma ray bursts, { Nucl. Phys. (Proc. Supl)} {\bf 91} (2000)
494-500 ; Gamma ray bursts, cosmic rays and neutrinos, { Nucl. Phys. (Proc. Supl)}
{\bf 87} (2000) 345-354.

\bibitem{VIETRI} M. Vietri, Ultrahigh-energy neutrinos from gamma ray bursts,
 { Phys. Rev. Letts.} {\bf 80}(1998) 3690-3693.

\bibitem{ROY} M. Roy, H.J. Crawford and A. Trattner,  The prediction and
detection of UHE Neutrino Bursts, [astro-ph/9903231].

\bibitem{cline} D.B. Cline and F.W. Stecker, Exploring the ultrahigh
energy neutrino universe, [astro-ph/0003459].

\bibitem{halzen} F. Halzen, High-energy neutrino astronomy,
 { Phys. Rep.} {\bf 333} (2000) 349-364.

\bibitem{URRU1} J. Alfaro, H.A. Morales-T\'ecotl and L.F. Urrutia, Quantum gravity corrections
to neutrino propagation, {Phys. Rev. Letts.} {\bf 84}(2000)2318, [gr-qc/9909079].

\bibitem{ELLISFERM} J. Ellis, N.E. Mavromatos, D.V. Nanopoulos and G.
Volkov, Gravitational recoil effects of fermion propagation in spacetime foam, {Gen.
Rel. Grav.} {\bf 32} (2000) 1777-1798, [gr-qc/9911055].

\bibitem{volumeop}
C. Rovelli and L. Smolin, Discreteness of area and volume in quantum gravity, { Nucl.
Phys.} {\bf B442} (1995) 593-622; Erratum-ibid {\bf B456} (1995) 753. A. Ashtekar and
J. Lewandowsky, Quantum theory of geometry I: Area operators, { Class. Quant. Grav.}
{\bf 14} (1997) A55-A82; Quantum theory of geometry II: Volume operators, { Adv.
Theor. Math. Phys.} {\bf 1} (1998) 388-429. A. Ashtekar, A. Corichi and J. Zapata,
Quantum theory of geometry III: Noncommutativity of Riemannian structures, { Class.
quant. Grav.} {\bf 15} (1998) 2955-2972.

\bibitem{CanonicalBlackHole}
C. Rovelli, Loop quantum gravity and black hole physics, Helv. Phys. Acta {\bf 69}
(1996) 582-611 [gr-qc/9608032]. C. Rovelli, Black hole entropy from loop quantum
gravity, Phys. Rev. Lett.{\bf 77} (1996) 3288-3291,1996 [gr-qc/9603063]. A. Ashtekar,
J. Baez, A. Corichi and K. Krasnov, Quantum geometry and black hole entropy, Phys.
Rev. Lett. {\bf 80} (1998) 904-907 [gr-qc/9710007].

\bibitem{canonicalnosingular}
M. Bojowald, Absence of singularity in loop quantum cosmology, Phys. Rev. Lett. {\bf
86} (2001) 5227-5230 [gr-qc/0102069].

\bibitem{RROV} For a recent review see for example C. Rovelli, Loop quantum gravity,
Livings Reviews Vol 1, 1998-1, http://www.livingreviews.org /Ar\-ti\-cles. See also
R. Gambini and J. Pullin, Loops, Knots, Gauge Theories and Quantum Gravity, Cambridge
University Press, Cambridge UK 1996.

\bibitem{Ash} A. Ashtekar, New variables for classical and quantum gravity,
{ Phys. Rev. Letts.} {\bf 57} (1987) {2244}. C. Beetle and A. Corichi, { Bibliography
of publications related to classical and quantum gravity in terms of connections and
loop variables } [gr-qc/9703044].

\bibitem{qdiffgauge}
A. Ashtekar, J. Lewandowski, D. Marolf, J. Mourao and T. Thiemann, Quantization of
diffeomorphism invariant theories of connections with local degrees of freedom, Jour.
Math. Phys. {\bf 36} (1995) 6456-6493 [gr-qc/9504018].


\bibitem{Barbero} F. Barbero, Real Ashtekar variables for Lorentzian signature space
times, {Phys. Rev. } {\bf D51} (1995) 5507-5510; Reality conditions and Ashtekar
variables: a different perspective, { Phys. Rev. } {\bf D51} (1995) {5498-5506}.


\bibitem{WICK} T. Thiemann, Reality conditions inducing transforms
for quantum gauge field theory and quantum gravity, { Class. Quant. Grav.} {\bf 13}
(1996) 1383-1404 [gr-qc/9511057]. Ashtekar, A generalized Wick transform for gravity,
{ Phys. Rev.} {\bf D53} (1996) 2865-2869 [gr-qc/9511083]. A. Ashtekar, J.
Lewandowski, D. Marolf, J. Mourao, T. Thiemann, Coherent state transforms for spaces
of connections, { J. Funct. Anal.} {\bf 135} (1996) 519-551 [gr-qc/9412014].

\bibitem{REALITY} H.A. Morales-T\'ecotl, L.F. Urrutia, J.D. Vergara, Reality conditions
for Ashtekar variables as Dirac constraints, { Class. Quant. Grav.} {\bf 13} (1996)
2933-2940 [gr-qc/9607044]. M. Montesinos, H.A. Morales-Tecotl, L.F. Urrutia and J.D.
Vergara, Real sector of the nonminimally coupled scalar field to selfdual gravity, {
J. Math. Phys.} {\bf 40} (1999) 1504-1517 [gr-qc/9903043];  Complex canonical gravity
and reality constraints, { Gen. Rel. Grav.} {\bf 31} (1999) 719-723.

\bibitem{Thiemann} T. Thiemann, QSD5: Quantum gravity as the natural regulator
of matter quantum field theories,  {\em Class. Quan. Grav.} {\bf 15} (1998) 1281-1314;
{Quantum Spin Dynamics (QSD) }, {\em Class. Quant. Grav.} {\bf 15} (1998) 839-873.

\bibitem{weave} A. Ashtekar, C. Rovelli and L. Smolin, Weaving quantum geometries
with quantum threads, {\em Phys. Rev. Letts.} {\bf 69}(1992) 237-240,
[hep-th/9203079]. J. Zegwaard, The weaving of curved geometries, {\em Phys. Lett.}
{\bf B300} (1993) 217-222 [hep-th/9210033]. R. Borissov, Weave states for plane
gravitational waves, {\em Phys. Rev.} {\bf D49} (1994) 923-929. J. Iwasaki and C.
Rovelli, Gravitons as embroidery on the weave, {\it Int. J. Mod. Phys.} {\bf D\ 1}
(1993) 533-557; Gravitons from loops: nonperturbative loop space quantum gravity
contains the graviton physics information, {\em Class. and Quant. Grav.} {\bf 11}
(1994) 1653-1676. J. Iwasaki, {\it Basis states for gravitons in nonperturbative loop
representation space} [gr-qc/9807013].

\bibitem{intersecting} B. Brugmann, R. Gambini and J. Pullin,  Knot
invariants as nondegenerate quantum geometries, {\em Phys. Rev. Letts.} {\bf 68}
(1992) 431-434.



\bibitem{twocircles}
N. Grot and C. Rovelli, Weave states in loop quantum gravity, {\em Gen. Rel. Grav.}
{\bf 29} (1997) 1039-1048.

\bibitem{gaussweave}
A. Corichi and J.M. Reyes, A Gaussian weave for kinematical loop quantum gravity,
Int. J. Mod. Phys. D {\bf 10} (2001) 325-338 [gr-qc/0006067].

\bibitem{aei} T. Thiemann, Gauge field theory coherent states (GCS) I: General properties,
{\em Class. Quant. Grav.} {\bf 18} (2001) 2025-2064, [hep-th/0005233]; T. Thiemann and
O. Winkler, Gauge field theory coherent states (GCS) II: Peakedness properties, {\em
Class. Quant. Grav.} {\bf 18} (2001) 2561-2636, [hep-th/0005237]. T. Thiemann and O.
Winkler, Gauge field theory coherent states (GCS): III. Ehrenfest theorems
[hep-th/0005234]; Gauge field theory coherent states (GCS): IV. Infinite tensor
product and thermodynamical limit [hep-th/0005235]. H. Sahlmann, T. Thiemann, O.
Winkler, { Coherent States for Canonical Quantum General Relativity and the Infinite
Tensor Product Extension} [gr-qc/0102038].

\bibitem{statg} A. Ashtekar and L. Bombelli, Statistical geometry of
quantum spin networks: flat space, in preparation;
L. Bombelli, Statistical geometry of random weave states, [gr-qc/0101080].

\bibitem{HEITLER} For the analogous situation in  electrodynamics see for example W. Heitler, {\it Quantum Theory of Radiation}, 3$^{\rm rd}$ edition, Clarendon Press, Oxford, England (1954).

\bibitem{Landau} L.D. Landau and E.M. Lifshitz, Quantum Electrodynamics, Pergamon Press, 1967,
page 585.

\bibitem{jacobson}
T. Jacobson and D. Mattingly, Generally covariant
model of a scalar field with high frequency dispersion and the cosmological horizon
problem, Phys. Rev. {\bf D63} (2001) 041502(R);
T. Jacobson, Lorentz violation and Hawking radiation, [gr-qc/0110079].

\bibitem{madzap}
M. Varadarajan and  J.A. Zapata, A proposal for analyzing the classical limit of
kinematic loop quantum gravity, Class. Quant. Grav. {\bf 17} (2000) 4085-4110
[gr-qc/0001040].

\bibitem{FockU1}
M. Varadarajan, Fock representations from U(1) holonomy algebras, Phys. Rev. D {\bf
61} (2000) 104001 [gr-qc/0001050]; Photons from quantized electric flux
representations [gr-qc/0104051].

\bibitem{Fockpolymer}
A. Ashtekar and J. Lewandowski, Relation between polymer and Fock excitations,
Class.Quant.Grav. {\bf 18}L117-L128(2001), [gr-qc/0107043].

\bibitem{AMC4} G. Amelino-Camelia, Space-time quantum solves three experimental paradoxes, [gr-qc/0107086].

\bibitem{ALF} J. Alfaro and G. Palma, Loop Quantum Gravity corrections and Cosmic Rays decays, [hep-th/0111176].

\bibitem{kostelecky} D. Colladay and V.A. Kostelecky, Lorentz-violating extension of the Standard Model, Phys. Rev. {\bf D58} (1998) 116002 [hep-ph/9809521]. For a recent review see for example V.A. Kostelecky, Topics in Lorentz and CPT violation, [hep-ph/0104227] and references therein.

\bibitem{tritium}
J.M. Carmona and  J.L. Cort\'es, Testing Lorentz invariance in the Tritium Beta decay
anomaly, Phys. Lett. B {\bf 494}(2000)75-80 [hep-ph/0007057]; Infrared and
ultraviolet cutoffs of quantum field theory [hep-th/0012028].

\bibitem{liberati}
S. Liberati, T. Jacobson and  D. Mattingly,
High-energy constraints on Lorentz symmetry violations,
[hep-ph/0110094]; T. Jacobson, Stefano Liberati and D. Mattingly,
TeV astrophysics constraints on Planck scale Lorentz violation, [hep-ph/0112207].

\bibitem{gleiser}
R.J. Gleiser and C.N. Kozameh, Astrophysical limits on quantum gravity motivated birefringence,
Phys. Rev. {\bf D64}083007 (2001), [gr-qc/0102093].

\end{thebibliography}
\end{document}